\documentclass[aps,nofootinbib,onecolumn]{revtex4}
\usepackage[pdftex]{graphicx}
\usepackage{epstopdf}
\usepackage{subfig}
\usepackage{amssymb}
\usepackage{amsmath}
\usepackage{bm}
\usepackage{slashbox}
\usepackage{mathtools}
\usepackage{xcolor}
\usepackage{amssymb}

\begin{document}

\title{Nonminimal global monopole}
%\date{\today}
\author{Thiago R. P. Caram\^es}
\email{trpcarames@id.uff.br}
\affiliation{Departamento de Ci\^encias Exatas, Biol\'ogicas e da Terra, Universidade Federal Fluminense (UFF),\\
Santo Ant\^onio de P\'adua, Rio de Janeiro 28470-000, Brazil}


\begin{abstract} 
In this paper, we investigate the gravitational effects of a global monopole that couples nonminimally to gravity. Considering a coupling parameter of arbitrary strength, we have obtained an analytical solution for the field equations of the model in the asymptotic region outside the monopole’s core, thus generalizing previous results. Within the weak coupling regime, we have also analyzed how this matter-geometry interplay affects some physical properties of the monopole, such as its radius and the mass enclosed in its core, which are explicitly computed and expressed in terms of the parameter of the theory. Using the Hahari-Loustó toy model, we have also found that the modification of general relativity (GR) may render a positive sign to the monopole mass, which means a global monopole able to exert attractive gravitational force on matter particles surrounding it. Next, we have studied some properties of a hairy black hole carrying such a global monopole charge. In this vein, we verified that the deviation from GR may provide such a spacetime with both event and Cauchy horizons, thus emulating a Reissner-Nordström framework.
Finally, we have performed a study of the geodesic motion of timelike particles around this black hole by setting the conditions for stable circular orbits. We also examined the gravitational bending experienced by photons traveling through this geometry, taking into account finite-distance corrections on the deflection angle. Additionally, we make an order-of-magnitude estimate for this extra contribution considering the
light bending caused by  Sagittarius A*.
\end{abstract}

\maketitle
\noindent
\section{Introduction}
Topological defects are ubiquitous in condensed matter physics: magnetic systems, elastic solids and liquid crystals are some of the physical systems where one usually comes across different types of such interesting structures \cite{mermin,chuang,katanaev,kroner}. In all these cases, these defects arise as result of a symmetry breaking triggered by previous phase transitions underwent by the system \cite{kibble,vach}.  
The possible existence of these configurations in the cosmological realm is one of the most relevant predictions of the grand unified theories. Like their condensed matter counterparts, these cosmic objects would be consequence of processes of spontaneous symmetry breaking induced by the several phase transitions experienced by the early Universe \cite{vilenkin}. According to the Kibble mechanism \cite{kibble}, the type of symmetry that is broken down would determine the specific topological defect to be produced. For instance, the breaking of an $SO(3)$ symmetry gives rise to a pointlike topological defect called global monopole, whose gravitational effects were first analyzed in a seminal paper published in 1989 by Barriola and Vilenkin \cite{barriola}. In that article, the authors discussed many remarkable features related to the gravitating global monopole in the context of General Relativity (GR). One of these aspects that we could mention is the solid deficit angle that characterizes the spacetime surrounding this defect. On one hand, they have shown that for a negligible monopole mass such a solid angle deficit precludes the global monopole to exert Newtonian force on timelike particles moving around it. On the other hand, it is shown that such a nontrivial topology does affect any light geodesics traveling through the monopole's vicinity. Just after this pioneer work, Hahari and Loust\'o went further by discussing possible impacts that gravity may have on some physical properties that shape the monopole's inner structure \cite{lousto}, like its mass and the size of its core. In this regard, they have obtained by both analytical and numerical techniques the radius of the monopole's core and the mass contained in it. One noteworthy point of their results is certainly the negative sign for such a mass, which was interpreted as a repulsive gravitational potential emerging around the defect, as a result of the de Sitter-like nature of its interior solution. More discussion concerning this monopole's mass issue may be found in Refs. \cite{nucamendi0,nucamendi}.
Global monopoles were also extensively studied within other models of gravity, with different implications of its gravitational effects being explored  \cite{romero,carames1,carames2,man,cheng,jpmg,valdir,carames3,lamb,PB}. 

%The GR was surely one of the most fascinating scientific endeavours of the last century. It has been revolutionized our comprehension of space, time and gravity and since it was presented by Albert Einstein in 1915, the sucessive observational confirmations of their main predictions helped to atest its robusteness as a physical theory. The most recent of them was the gravitational waves detection , that have opened the doors for a new physics and paved a promising road to be explored by the research in gravity in the next decades. 

Despite all the recognized achievements of GR in describing properly the majority of the known gravitational phenomena, there are still some open issues putting the Einstein theory at a challenging crossroad. The appearance of singularities in some of the most important GR solutions, notably, the cosmological and the black holes ones, may be indicating that this theory needs to be modified to some extent, especially when one deals with high energy regimes. Some authors claim that an accurate answer for this pathological property could emerge within a quantum representation of Einstein gravity (see \cite{bosma} and the references therein). However, the implementation of this quantum description encounters serious difficulties due to the nonrenormalizable character of GR \cite{zee}. 

Furthermore, the late dynamics of the observed large scale Universe shows a current speedup of the cosmic expansion. The standard cosmological model, ($\Lambda$ cold dark matter), which has GR as its underlying gravitational theory, is just capable to explain such a behavior by considering a cosmological constant ($\Lambda$) as part of the Einstein equations. In this case, such a new term would behave as a negative pressure fluid that would produce at large scales the repulsive gravity necessary to account for the observed cosmic acceleration. Nonetheless, the presence of cosmological constant in the theory brings an inconsistency between cosmological observations and the predictions of the quantum field theory when one gives to $\Lambda$ the standard interpretation of a vacuum energy density. This is the so-called ``cosmological constant problem" \cite{zeld,zeld1,sola} . 

For these reasons, we have seen in the recent decades an increased search for alternative theories of gravity, each of them proposing different paths to modify the GR. In the most common proposals, the existence of scalar degrees of freedom is assumed through the introduction of some dynamical scalar field in the gravitational action. This happens with the scalar-tensor models, mainly the wide class of Horndeski theories and, consequently, some of their typical particular cases, such as Brans-Dicke and $f(R)$ gravity, for example \cite{horndeski}. However, other modified theories follow a bolder recipe, by assuming that some premises of GR may be relaxed. This is the case of the gravitational theory under consideration in this work. It is well known that the principle of minimal matter-curvature coupling is one of GR's cornerstones, which has as one of its main consequences the covariant conservation of the energy-momentum tensor, which also means that the free matter particles follow geodesic paths. Consequently, it is clear that the breaking of the minimal coupling assumption also leads to violation of the equivalence principle \cite{faraoni,velten}. 

In this paper we work within the nonminimally coupled gravity theory proposed in \cite{orfeu}. As the literature shows us, this theory was widely explored in different branches. It was proposed as a potential way out for the dark matter problem \cite{orfeuDM,orfeuDM-1}, to model stellar equilibrium \cite{fR-matter}, and as a possible ingredient responsible for the current cosmic acceleration \cite{orfeuCA}; also produced interesting results in the context of the cosmological perturbations \cite{orfeu-pert}, gravitational waves \cite{orfeu-GW}, inflationary scenarios \cite{gomes-infl} and energy conditions \cite{sequeira}. We must also mention the studies of observational constraints coming from these theories at the Solar System \cite{orfeu-solar, march} and cosmological background levels \cite{an}, as well as the bounds imposed by ocean experiments \cite{orfeu-ocean}.   
 
In the present article we consider a new model of gravitating global monopole in which the monopole couples nonminimally to the spacetime curvature. This nonminimal coupling is modeled through the presence of a mixing between matter and geometric sectors. This mixing term is implemented in the generalized gravitational action by means of an arbitrary function of the Ricci scalar. The form of this function is chosen in such a manner that we are left with the simplest version of this nonminimally coupled gravity. Moreover, we also discuss other criteria to justify our choice on the basis of what has been usually adopted in the literature. It is shown that this novel aspect brings new contributions to previous studies within the global monopole physics as we shall see throughout this work.   

This paper is organized as follows. In Sec. II, we introduce the basic ingredients of a global monopole model within a nonminimally coupled gravity. In Sec. III, we present an analytical solution for the gravitational field of this nonminimal global monopole within the outside-the-core approximation, first assuming an arbitrary magnitude for the coupling parameter, then by considering the weak coupling case, for which the solution takes a much simpler form. In the section IV, we investigate the inner solution of the global monopole, analyzing how the nonminimal matter-curvature coupling affects its inner structure, in particular its mass and the size of its core. In Sec. V, we study the gravitational field of a hypothetical black hole endowed with a nonminimal global monopole charge, by discussing the appearance of (extra) event and Cauchy horizons in such a spacetime. Next, in Sec. VI, we have performed a study of the geodesic motion of timelike particles, showing that the model admits the arising of both new stable and unstable orbits in this background. In Sec. VII, we have examined the consequences of the nonminimal coupling on the light bending phenomenon, where we have provided an order-of-magnitude estimate of the correction on the deflection angle. Finally, in Sec. VIII we summarize the main results of this work and leave our concluding remarks. 

\section{The model}
\subsection{The nonminimally coupled gravity}
A theory of gravity with nonminimal matter-curvature coupling may be modeled in different ways. We use here the one proposed in \cite{orfeu}, where the authors depart from the following gravitational action:
\begin{equation}
\label{action}
S=\int \left\{ \frac{1}{2}f_{1}(R)+\left[1+\alpha f_{2}(R)\right]{\cal L}_{m}\right\}\sqrt{-g} d^{4}x,
\end{equation}
where ${\cal L}_{m}$ is the matter Lagrangian, whereas $f_1(R)$ and $f_2(R)$ are arbitrary functions of the Ricci scalar. The parameter $\alpha$ encodes the information of how strong is the nonminimal coupling. 

The variation of (\ref{action}) with respect to the metric tensor provides the set of field equations below
\begin{equation}
\label{FEQs} 
 \left(F_1+2\alpha F_2 {\cal L}_{m}\right)R_{\mu\nu}-\frac{1}{2}f_{1}g_{\mu\nu}=\left(\nabla_{\mu}\nabla_{\nu}-g_{\mu\nu}\Box\right)\left(F_{1}+2\alpha F_{2} {\cal L}_{m}\right)+(1+\alpha f_2)T_{\mu\nu},
\end{equation}
where one defines $F_{i}\equiv\frac{df_i}{dR}$. Notice that the GR case is recovered when $f_1(R)=\frac{R}{\kappa}$ and $\alpha=0$. Furthermore, it is obvious that the vacuum solutions of both theories must be the same at the limit $f_1(R)=\frac{R}{\kappa}$, since in this situation there is no matter-curvature coupling and this modified gravity coincides with GR. 
In this theory, the energy-momentum does not respect the usual conservation law; it is instead subject to a nontrivial condition given by
\begin{equation}
\label{conslaw}
\nabla^{\mu}T_{\mu\nu}=\frac{\alpha F_{2}}{1+\alpha f_{2}}\left[g_{\mu\nu}{\cal L}_{m}-T_{\mu\nu}\right]\nabla^{\mu}R,
\end{equation}
with the extra term on the right-hand side describing the energy-momentum exchange between matter and curvature. 
%In \cite{orfeu} we have seen that such a conservation law is obtained by means of the Bianchi identities, $\nabla^{\mu}G_{\mu\nu}=0$ along with the relation $(\Box\nabla_{\nu}-\nabla_{\nu}\Box)F_{i}=R_{\mu\nu}\nabla^{\mu}F_{i}$.
In this study, we shall assume the simplest nonminimally coupled gravity model where the functions $f_{1}(R)$ and $f_{2}(R)$ are given by 
\begin{equation}
\label{fns}
f_1(R)=\frac{R}{\kappa}\;\;\textrm{and}\;\;\;f_2(R)=R,   
\end{equation}
where $\kappa=\frac{8 \pi G}{c^4}$ is the gravitational coupling constant. Notice that the model above resembles GR, except for the nonminimal interaction between the matter and geometric sectors represented by the factor $\alpha R {\cal L}_{m}$. The choice (\ref{fns}) is particularly interesting since it enhances such a nonminimal coupling: by leaving $f_{1}(R)$ as the usual Einstein-Hilbert Lagrangian, we ensure that all the new effects arising in this scenario are an exclusive result of the non-trivial interplay between matter and geometry. 
There is, however, a discussion in the literature about the suitable form for $f_{2}(R)$. As we can see in \cite{an}, in the cosmological domain the proper ansatz is a power law $f_2(R) \propto R^{n}$ with $n<0$, as it fits very well observational datasets, describing correctly the evolution of the observed Universe. On the other hand, the linear choice (\ref{fns}) reveals to be appropriate on small scales or at high matter density contexts. This includes several gravitational phenomena comprising geophysical, Solar System and astrophysical scales. 
%In \cite{fisher} the authors go beyond by resorting to the same choice (\ref{fns}) to investigate potential bounds on the nonminimal coupling parameter coming from nuclear physics. 
Since in this study we are interested in phenomena that are relevant at most on the astrophysical environment, the choice $f_{2}(R)=R$ is adequate to fulfill the aims of this work.

A possible experimental limit on the linear models $f_{2}(R)=R$ has been considered in Ref. \cite{fisher}, where the authors have used nuclear physics to impose on the parameter $\alpha$ a stringent upper bound of $|\alpha|<5\times10^{-12}\;\textrm{m}^2$. However, as it is argued in \cite{march}, this result  stems from a strong suppression at the high density environment of the nuclear physics scale as a consequence of a screening mechanism. Since we are not dealing with systems characterized by such high matter densities, like those ones verified in the interior of astrophysical bodies, for example, this upper bound will not be taken into account in our study. 

Using (\ref{fns}) in (\ref{FEQs}) we arrive at 
\begin{equation}
\label{FEQs1} 
(1+2\alpha \kappa {\cal L}_{m})R_{\mu\nu}-\frac{1}{2}g_{\mu\nu}R=2\alpha \kappa \left(\nabla_{\mu}\nabla_{\mu}-g_{\mu\nu}\Box\right){\cal L}_{m}+\left(1+\alpha R\right)\kappa T_{\mu\nu}.
\end{equation}
Additionally, it is useful to obtain the trace of the field equations (\ref{FEQs1}),
\begin{equation}
\label{trace} 
R=\frac{-\kappa T + 6\alpha \kappa \Box {\cal L}_{m}}{1-2\alpha \kappa {\cal L}_{m}+\alpha \kappa T}, 
\end{equation}
from which the GR relation $R=-\kappa T$ is promptly recovered when $\alpha=0$. 

%As it is discussed in \cite{sequeira}, with the presence of such %a nonminimal coupling it is possible to define of an effective %gravitational coupling which is proportional to the term $\alpha %F_{2}$.   Throughout this work we shall use only positive values %for the nonminimal coupling parameter, $\alpha$.  

\subsection{The global monopole spacetime}

We are interested in specializing the gravitational equations above to the case of a gravitating global monopole. It is known that the Lagrangian density associated with this configuration is
\begin{equation}
\label{GMlag}
{\cal L}_{m}=-\frac{1}{2}\partial_{\mu}\phi^{a} \partial^{\mu}\phi^{a}-\frac{1}{4} \lambda (\phi^{a}\phi^{a}-\eta^2)^2\ ,
\end{equation}
which shows clearly the breaking of the $SO(3)$ group to $U(1)$ that leads to the formation of the monopole. The parameter $\lambda$ is the self-coupling constant of the Higgs field, whereas $\eta$  means the energy scale of the symmetry breaking. Moreover, the Higgs field $\phi^a$ consists of an isotriplet of scalar fields whose form is given by the so-called hedgehog ansatz, 
\begin{equation}
\label{campo}
\phi^{a}=\eta h(r) \hat{x}^{a}\ ,
\end{equation}
with the index $a=1,2,3$ and
\begin{equation}
\label{versor}
\hat{x}^{a}=\{\sin \theta \cos \varphi, \sin \theta \sin \varphi, \cos \theta \}.
\end{equation}
Additionally, the radial function $h(r)$ is subject to the following boundary conditions:
\begin{equation}
\label{BCs}
h(0)=0,\;\;\;\; h(r\rightarrow \infty)=1.
\end{equation}

Given the symmetry obeyed by a global monopole, let us analyze it by means of the spherically symmetric line element below,
\begin{equation}
\label{metric} 
ds^2=-B(r)dt^2+A(r)dr^2+r^2d\theta^2+r^2 \sin^2 \theta d\varphi^2. 
\end{equation}
Using (\ref{campo}) and (\ref{metric}) in (\ref{GMlag}) we obtain
\begin{equation}
\label{lag}
{\cal L}_{m}=-\frac{1}{2}\frac{\eta^2h'^2}{A}-\frac{\eta^2h^2}{r^2}-\frac{\lambda \eta^4}{4}(h^2-1)^2.
\end{equation}
Additionally, it is well known that the energy-momentum tensor is defined in terms of the matter Lagrangian as 
\begin{equation}
\label{emt}
T_{\mu\nu}=-\frac{2}{\sqrt{-g}}\frac{\delta \left(\sqrt{-g}{\cal L}_{m}\right)}{\delta g^{\mu\nu}}. 
\end{equation}
Using the equations given above we may find the nonvanishing components of the energy-momentum tensor as follows:
\begin{eqnarray}
\label{emt1}
T^{0}_{\;\;0}=-\eta^2 \left[\frac{h'^2}{2A}+\frac{h^2}{r^2}+\frac{\lambda}{4}\eta^2\left(h^2-1\right)^2\right] ,\nonumber\\
T^{1}_{\;\;1}=-\eta^2 \left[-\frac{h'^2}{2A}+\frac{h^2}{r^2}+\frac{\lambda}{4}\eta^2\left(h^2-1\right)^2\right] ,\\
T^{2}_{\;\;2}=T^{3}_{\;\;3}=-\eta^2 \left[\frac{h'^2}{2A}+\frac{\lambda}{4}\eta^2\left(h^2-1\right)^2\right] ,\nonumber
\end{eqnarray}
as well as its trace
\begin{equation}
\label{TT}
T=-\frac{\eta^2h'^{2}}{A}-\frac{2\eta^2 h^2}{r^2}-\lambda \eta^4(h^2-1)^2.  
\end{equation}

From (\ref{conslaw}) it is possible to determine the dynamical equation governing the Higgs field behavior for the background metric  (\ref{metric}), which is
\begin{equation}
\label{phi}
h''+\left[\frac{2}{r}+\frac{1}{2}\left(\frac{B'}{B}-\frac{A'}{A}\right)\right]h'-\frac{2Ah}{r^2}-\lambda \eta^2 A h\left(h^2-1\right)=\frac{\alpha R' h'}{1+\alpha R}.
\end{equation}
%For the weak coupling approximation the expression above becomes
%\begin{equation}
%\label{phi1}
%h''+\left[\frac{2}{r}+\frac{1}{2}\left(\frac{B'}{B}-\frac{A'}{A}\right)\right]h'-\frac{2Ah}{r^2}-\lambda \eta^2 A h\left(h^2-1\right)=\alpha R' h'.\end{equation}

Outside the monopole's core the Higgs field behaves as $h\approx1$, which leads to a quite simpler form for the components (\ref{emt1}),  
\begin{equation}
\label{emt2}
T_{\mu}^{\nu}\approx\textrm{diag}\left(-\frac{\eta^2}{r^2},-\frac{\eta^2}{r^2},0,0\right)\ ,
\end{equation}
whereas the Lagrangian ${\cal L}_{m}$ and the trace $T$ are reduced to
\begin{equation}
\label{lm1} 
 {\cal L}_{m}\approx-\frac{\eta^2}{r^2}
\end{equation}
and
\begin{equation}
\label{tt}
T\approx-\frac{2\eta^2}{r^2},
\end{equation}
respectively. 

The outside-the-core assumption refers to a region far from the defect's lump, where the Higgs field reaches its vacuum expectation value and is convenient for the purpose of this work for a couple of reasons. First of all, it is the regime in which the Barriola-Vilenkin solution has been obtained, which makes this choice useful for a suitable comparison between the standard global monopole and the nonminimal one under consideration in this study. Second, due to the smallness of the monopole's width it is expected that the condition $h\approx1$ applies for most of the physically relevant situations, especially those ones in the astrophysical context.   
%we expect that such   in the astrophyical context,    when using the it is easy to check that by using (\ref{emt2}) and (\ref{lm1}) into (\ref{conslaw}) we achieve the usual conservation law of the GR, namely, $\nabla^{\mu}T_{\mu\nu}=0$, even for a non-constant $f_{2}(R)$. This means that we have a particular case in which particles moving through the global monopole spacetime will follow geodesics paths even for a non-zero mixing between matter and geometry in the gravitational action.

Another useful quantity is $\Box {\cal L}_{m}$ which can be computed from (\ref{metric}) and (\ref{lm1}) giving
\begin{equation}
\label{BoxL} 
\Box {\cal L}_{m}=-\frac{2\eta^2}{Ar^4}+\frac{1}{A}\left(\frac{B'}{B}-\frac{A'}{A}\right)\frac{\eta^2}{r^3}. 
\end{equation} 

In \cite{barriola} the approximation (\ref{emt2}) has been used to first explore the gravitational effects of a global monopole. The obtained solution is given by the following metric functions
\begin{equation}
\label{bv}
B(r)=A(r)^{-1}=1-\kappa \eta^2-2GM/r\ .
\end{equation}
In that study, the authors offered a twofold interpretation for the integration constant $M$. The first one considers it as the mass of the monopole's core itself, whereas the second one takes $M$ as the mass of a black hole that swallowed a global monopole, thus inheriting the defect's charge, $\kappa \eta^2$. In their work, Barriola and Vilenkin discuss only the first possibility, although they eventually throw away $M$, arguing that it is negligible on the astrophysical scale. Thereafter, a change of variable in the solution (\ref{bv}) turns the monopole's charge into the so-called solid deficit angle term,
\begin{equation}
\label{BV-metric}
ds^2=-dt^2+dr^2+(1-\kappa \eta^2)r^2d\Omega^2.
\end{equation}

\section{Solving the field equations}
Let us now write the main geometric quantities necessary to obtain the field equations in their explicit form. We are dealing with a spherically symmetric problem, in which the line element (\ref{metric}) is taken as a starting point. For such a metric, the nonzero Christoffel symbols are the following
\begin{eqnarray}
\label{CS}
\Gamma^0_{01} &=& \frac{B'}{2B},\;\;\;\;\;\; \quad \Gamma^{1}_{00} = \frac{B'}{2A}, \;\; \quad
\Gamma^{1}_{11} = \frac{A'}{2A},\nonumber\\ \Gamma^{2}_{12} &=& \Gamma^{3}_{13} = \frac{1}{r}, \quad \Gamma ^{3}_{32} = \cot\theta, \quad
\Gamma^{1}_{22} = - \frac{r}{A}, \\ \Gamma^{1}_{33} &=& - \frac{r}{A}\sin^{2}\theta , \quad \Gamma^{3}_{23} =-\sin\theta\cos\theta.\nonumber
\end{eqnarray}
On the other hand, the nonzero components of the Ricci tensor associated with the metric (\ref{metric}) are
\begin{eqnarray}
\label{RTensor}
R^{0}_{\;0}=\frac{1}{2}\left[\frac{B'^2}{2B^2A}+\frac{B'A'}{2BA^2}-\frac{B''}{BA}-\frac{2B'}{BAr}\right] ,\nonumber\\
R^{1}_{\;1}=\frac{1}{2}\left[\frac{B'A'}{2BA^2}+\frac{2A'}{rA^2}-\frac{B''}{BA}+\frac{B'^2}{2B^2A}\right] ,\\
R^{2}_{\;2}=R^{3}_{\;3}=\frac{A'}{2rA^2}-\frac{B'}{2rBA}-\frac{1}{r^2A}+\frac{1}{r^2}. \nonumber
\end{eqnarray}
We may express the scalar curvature in terms of the matter configuration by means of (\ref{trace}). From this relation we find   
\begin{equation}
\label{Rscalar}
R=\frac{2\kappa \eta^2}{r^2}+\frac{12\alpha \kappa^2 \eta^4}{r^4}-\frac{6\alpha \kappa \eta^2}{A r^3}\left(\frac{B'}{B}-\frac{A'}{A}\right),
\end{equation}
where we have used the quantities (\ref{lm1})--(\ref{BoxL}). 

With the equations above at hand, we are now ready to construct the field equations for a global monopole in nonminimally coupled gravity. Since we are searching for a solution very far from the monopole's core, we will substitute (\ref{lm1}) and (\ref{tt}) into the Eq. (\ref{FEQs1}). This provides the following set of equations of motion
\begin{align}
-&\left(1-\frac{2\alpha \kappa \eta^2}{r^2}\right) R_{0}^{0}+\frac{R}{2}=-\frac{2\alpha \kappa \eta^2}{r^3}\frac{B'}{AB}+2\alpha \kappa \Box {\cal L}_{m}+\left(1+\alpha R\right)\frac{\kappa \eta^2}{r^2}, \label{eom1}\\
&\left(1-\frac{2\alpha \kappa \eta^2}{r^2}\right) R_{1}^{1}-\frac{R}{2}=-\frac{12\alpha \kappa \eta^2}{Ar^4}-\frac{2\alpha \kappa \eta^2}{r^3}\frac{A'}{A^2}-2\alpha \kappa \Box {\cal L}_{m}-(1+\alpha R)\frac{\kappa \eta^2}{r^2},\label{eom2}\\
&\left(1-\frac{2\alpha \kappa \eta^2}{r^2}\right) R_{2}^{2}-\frac{R}{2}=\frac{4\alpha \kappa \eta^2}{Ar^4}-2\alpha \kappa \Box {\cal L}_{m}.\label{eom3}
\end{align}

In what follows let us obtain the solution for the system of nonlinear equations above. Our first step is to combine (\ref{eom1}) and (\ref{eom2}) to get the relation 
\begin{equation}
A(r)B(r)=c_0 e^{\left(\frac{6\alpha \kappa \eta^2}{r^2}\right)},\nonumber
\end{equation}
or
\begin{equation}
\label{AB}
A(r)B(r)=e^{\left(\frac{6\alpha \kappa \eta^2}{r^2}\right)},
\end{equation}
by choosing the integration constant $c_0=1$.

Equation (\ref{eom3}) provides the second independent field equation, which is given by the first order ordinary differential equation below,    
\begin{equation}
\label{2ndFE} 
(r^2-12\alpha \Delta)f'(r)+\frac{18\alpha \Delta}{r}\left(1+\frac{4\alpha \Delta}{r^2}\right)f(r)+2\alpha \Delta - (1-\Delta)r^2=0, 
\end{equation}
where the following definitions were used: 
\begin{align}
\Delta\equiv \kappa \eta^2,\label{delta}
\end{align}
and 
\begin{equation}
\label{fr}
f(r)\equiv \frac{r}{A}. 
\end{equation}

Equation (\ref{2ndFE}) admits the following analytical solution: 
\begin{equation}
\label{Ar}
A(r)^{-1}=\left(1-\frac{12 \alpha \Delta}{r^2}\right)^{-1}\left[1-\Delta +\frac{c_{1}}{r}e^{-\frac{3\alpha \Delta}{r^2}}+\frac{\sqrt{3\pi \alpha \Delta}}{3r}e^{-\frac{3\alpha \Delta}{r^2}}\textrm{Erfi}\left(\frac{\sqrt{3\alpha \Delta}}{r}\right)\right].
\end{equation}
Having (\ref{Ar}) at hand, the metric function $B(r)$ can be obtained with the help of the relation (\ref{AB}), which gives
\begin{equation}
\label{Br}
B(r)=\left(1-\frac{12 \alpha \Delta}{r^2}\right)^{-1}e^{\frac{6\alpha \Delta}{r^2}} \left[1-\Delta +\frac{c_{1}}{r}e^{-\frac{3\alpha \Delta}{r^2}}+\frac{\sqrt{3\pi \alpha \Delta}}{3r}e^{-\frac{3\alpha \Delta}{r^2}}\textrm{Erfi}\left(\frac{\sqrt{3\alpha \Delta}}{r}\right)\right], 
\end{equation}
where $\textrm{Erfi}(z)\equiv\frac{2}{\sqrt{\pi}}\int_{0}^{z}e^{x^2}dx$ is the imaginary error function
\cite{notes}. Equations (\ref{Ar}) and (\ref{Br}) represent an entirely new result in the literature, which is the outside-the-core solution for a gravitating global monopole nonminimally coupled to gravity that is modeled by the choice (\ref{fns}). The integration constant present in this solution can be set to $c_1=-2GM$, in order to recover the Newtonian regime in the appropriate limit. This assumption makes possible to retrieve the traditional Barriola-Vilenkin solution (\ref{bv}) when the coupling parameter vanishes.  

It is worth mentioning the interesting works \cite{nucamendi1,nucamendi2}, in which the authors also investigate global monopoles that are nonminimally coupled to gravity. In these studies, the authors introduce by hand a specific nonminimal coupling in the form $\propto\phi^{a}\phi^{a}R$,  as an attempt to approach the Galactic rotation curves problem.

%or, due to the weak coupling assumption,
%\begin{equation}
%\label{AB}
%A(r)B(r)\approx 1+\frac{6\alpha \kappa \eta^2}{r^2},
%\end{equation}
%where we set the integration constant as $c_0=1$. 

\subsection{The weak coupling case}
In order to better interpret some consequences of the solution given by (\ref{Ar}) and (\ref{Br}), we shall consider from now on a tiny coupling between matter and gravity, so that $\alpha f_{2}(R)=\alpha R<1$, which enables us  to retain in the obtained solution only the first order terms in the parameter $\alpha$, with all its nonlinear contributions being neglected. Since $\alpha$ has unit of $L^{2}$, its smallness in the weak coupling context is defined in comparison with some typical length scale $l_{0}$ associated with the problem under analysis, so that in this limit the condition $\alpha<l_0^{2}$ holds. For global monopole physics, it is inevitable to regard the radius of the core, $\delta \sim (\sqrt{\lambda} \eta)^{-1}$, as such length scale. Whereas, when the global monopole shows up as a charge of a black hole of mass $M$, one expects that $l_{0}\sim GM$. In the present work, both cases will be investigated, so that in each one the assumed weak matter-curvature coupling will be determined with respect to these two characteristic length scales.

%It seems more reasonable to consider the previous bound on $\alpha$ established  
Thus, the weak coupling assumption implies
\begin{eqnarray}
\label{approx} 
&&\left(1-\frac{12 \alpha \Delta}{r^2}\right)^{-1}=1+\frac{12\alpha \Delta}{r^2}+O(\alpha^2),\\
&& e^{-\frac{3\alpha \Delta}{r^2}}=1-\frac{3\alpha \Delta}{r^2}+O(\alpha^2),\\
&& \textrm{Erfi}\left(\frac{\sqrt{3\alpha \Delta}}{r}\right)=\frac{2}{\sqrt{\pi}}\frac{\sqrt{3\alpha \Delta}}{r}+O(\alpha^{3/2}).
\end{eqnarray}
%the trace (\ref{trace}) of the field equations will be reduced to 
%\begin{equation}
%\label{trace1} 
 %R\approx -\kappa T + 2\alpha \kappa^2
 %T {\cal L}_{m}- \alpha \kappa^2 T^2+ 6\alpha \kappa \Box {\cal L}_{m}
%\end{equation}.
This leads to the approximated solution below 
\begin{equation}
\label{solA} 
A(r)^{-1}\approx 1-\Delta - \frac{2GM}{r}+\frac{2\alpha \Delta(7-6\Delta)}{r^2}-\frac{18\alpha \Delta GM}{r^3},
\end{equation}
and
\begin{equation}
\label{solB} 
 B(r)\approx 1-\Delta - \frac{2GM}{r}+\frac{2\alpha \Delta(10-9\Delta)}{r^2}-\frac{30\alpha \Delta GM}{r^3}.
\end{equation}

So, the weak coupling hypothesis turns the analytical solution given by (\ref{Ar}) and (\ref{Br}) into a more tractable form. Additionally, such an assumption is also useful to preserve the attractive character of gravity within this modified theory, since it helps us to ensure positiveness for the effective gravitational constant without any further restriction on the choice of $f_{1}(R)$ and $f_{2}(R)$ \cite{sequeira}. 
%As we can see in the Refs.    quite reasonable, since the nonminimally coupled gravity implies in a violation of the equivalence principle \cite{faraoni,velten}. So, it is expected that the parameter $\alpha$ is strongly restricted by solar system tests, that shall constrain it to small enough values. 

Before studying in more detail the solution represented by (\ref{solA}) and (\ref{solB}), it is convenient adopt some further approximations. It is expected that the scale of symmetry breaking is quite below the Planck scale. Since $G \sim 1/M_{pl}^{2}$, it is suitable to assume $G\eta^2<1$ \cite{vilenkin}. This allows us to retain only the leading terms in the parameter $\Delta=8\pi G \eta^2$. Taking this assumption into account, the metric functions $A(r)$ and $B(r)$ become
%Since $G=1/M_{pl}^{2}\sim 10^{-38}\;\textrm{GeV}^{-2}$ and the symmetry breaking scale $\eta \sim 10^{16}\;\textrm{GeV}$, we shall have $G\eta^2\sim 10^{-6}$
\begin{equation}
\label{solA1} 
A(r)^{-1}\approx1-\Delta - \frac{2GM}{r}+\frac{14 \alpha \Delta}{r^2}-\frac{18 \alpha \Delta GM}{r^3},
\end{equation}
and
\begin{equation}
\label{solB1}     
 B(r)\approx1-\Delta - \frac{2GM}{r}+\frac{20\alpha \Delta}{r^2}-\frac{30 \alpha \Delta GM}{r^3}.
\end{equation}
Using the solution above in (\ref{Rscalar}), we find the following approximated solution for the Ricci scalar:
\begin{equation}
\label{Rscalar1}
R(r)\approx\frac{2\Delta}{r^2}-\frac{24\alpha\Delta GM}{r^5}.
\end{equation}
As it was previously pointed out in Sec. II, in the global monopole model there are two possible interpretations for the mass term present in the solution above. In the next sections, we will explore different facets of both cases. 

%it is knwon that it is provided two possible interpretations for the mass term \cite{barriola}. The first possibility is that such a term is the mass enclosed in the monopole's core as seen by an observer very far from the defect. While the second one considers the $M$ term as the mass of a Schwarzschild black hole holding a global monopole charge, which could be formed from a static star that collapsed near the monopole. 
%Now, with the solution above we are ready to explore in this work different facets of the both interpretations. 

\section{On the inner structure of the monopole}
\subsection{Computing the mass and the core radius}
Here we are interested in analyzing the potential changes that nonminimally coupled gravity may cause on physical properties that constitute the inner structure of the monopole. In \cite{lousto} the authors use a simple analytical model to compute both the radius and the monopole's mass. In that simplified model,  they assume that the monopole configuration consists of a pure false vacuum inside the core and a true vacuum outside. In other words, this means that the Higgs field  behaves as follows:
\begin{equation}
\label{HLmodel}
h(r)=
\begin{cases}
0 \;\; \textrm{if}\; r<\delta\;,\\
1 \;\; \textrm{if}\; r>\delta\;, 
\end{cases}
\end{equation}
where $\delta$ denotes the radius of the monopole's core. The field equations (\ref{FEQs1}) for $r<\delta$ provide the de Sitter-like solution below\footnote{For convenience, in this section we have opted for using the original notation, namely, in terms of $\kappa$, $\eta$, $\alpha$, and $\lambda$.},
\begin{equation}
\label{dS}
B(r)_{in}=A(r)^{-1}_{in}=1-\frac{\kappa \lambda \eta^4}{6(2-\alpha \kappa \lambda \eta^4)}r^2.
\end{equation}
Notice that the term in brackets introduces in the solution above contributions of higher order in $\kappa \eta^2$. Then, because of our approximative assumptions these can be neglected, which will result in the known GR solution \cite{lousto} 
\begin{equation}
\label{inC}
B(r)_{in}=A(r)^{-1}_{in}=1-\frac{\kappa \lambda \eta^4}{12}r^2.
\end{equation}

The exterior solution is the one we have already obtained in the last section, given by Eqs. (\ref{solA}) and (\ref{solB}). However, before proceeding to the next step, some considerations are in order. Let us recall that, as seen in \cite{barriola}, the monopole's core mass is $M_{c}\sim \eta$, while the weak coupling approximation when applied to the gravitational field of an isolated global monopole would constrain the parameter of the theory to $\alpha \lesssim \eta^{-2}$. So, if $M$ is now the mass inside the monopole's core and $r>\delta$ (external metric), it is easy to verify that the exterior solution should respect
\begin{equation}
\frac{\alpha \kappa \eta^2 GM_{c}}{r^3} < (G \eta^2)^2
\end{equation}
which is negligible, since 
$\eta\ll M_{pl}$. Therefore, we can ignore the contribution of these terms in the metric components (\ref{solA}) and (\ref{solB}), so that the exterior solution gets reduced to
\begin{equation}
\label{solA2} 
A(r)^{-1}_{out}=1-\kappa \eta^2 - \frac{2G M_{c}}{r}+\frac{14\alpha \kappa \eta^2}{r^2},
\end{equation}
and
\begin{equation}
\label{solB2}     
 B(r)_{out}=1-\kappa \eta^2 - \frac{2GM_{c}}{r}+\frac{20\alpha \kappa \eta^2}{r^2}. 
\end{equation}    
The inner and outer solutions shall match at $r=\delta$, which means to assume continuity across the boundary for the metric and its first derivative with respect to $r$. Imposing these requirements to the component $g_{00}(r)=B(r)$, we have
\begin{equation}
B_{in}(r=\delta)=B_{out}(r=\delta)
\end{equation}
and
\begin{equation}
\left.\frac{dB_{in}(r)}{dr}\right|_{r=\delta}=\left.\frac{dB_{out}(r)}{dr}\right|_{r=\delta}. 
\end{equation}
These two conditions give rise to the set of algebraic equations below, 
\begin{eqnarray}
\label{set} 
&&\frac{\kappa \lambda \eta^4}{12}\delta^4-\kappa\eta^2 \delta^2 - 2GM_{c} \delta+20 \alpha \kappa \eta^2=0\nonumber\\
&&-\frac{\kappa \lambda \eta^4}{6}\delta^4-2GM_{c} \delta+40 \alpha \kappa \eta^2=0,
\end{eqnarray}
which can be solved for $\delta$ and $M_{c}$. For the core radius we find the following pair of solutions:
\begin{equation}
\label{dplus} 
\delta_{+}=\frac{\sqrt{2}}{\sqrt{\lambda}\eta}\sqrt{1+\sqrt{1+20\tilde{\alpha}}} 
\end{equation}
and
\begin{equation}
\label{dminus} 
\delta_{-}=\frac{\sqrt{2}}{\sqrt{\lambda}\eta}\sqrt{1-\sqrt{1+20\tilde{\alpha}}}, 
\end{equation}
where we define the dimensionless parameter $\tilde{\alpha}\equiv \alpha \lambda \eta^2$. This duplicity in the solution for the radius core is a novel aspect arising due to the nonminimally coupled gravity. As will be seen in this section, we shall impose to the parameter $\tilde{\alpha}$ a range that assures real values for (\ref{dplus}) and (\ref{dminus}). For $\tilde{\alpha}=0$ both $\delta$'s above coincide with the GR one,
\begin{equation}
\label{dgr} 
\left.\delta_{+}\right|_{\alpha=0}=\left.\delta_{-}\right|_{\alpha=0}= \delta_{GR}\equiv \frac{2}{\sqrt{\lambda} \eta}. 
\end{equation}

Let us simplify the notation by adopting the parameter $\epsilon=+1\;\textrm{or}-1$ which carries the signs that distinguish (\ref{dplus}) and (\ref{dminus}). Additionally, let us write it in terms of $\delta_{GR}$,
\begin{equation}
\label{dmain} 
\delta=\delta_{GR}\sqrt{\frac{1}{2}+\frac{\epsilon}{2}\sqrt{1+20\tilde{\alpha}}}\;.
\end{equation}
The expression above tells us that with respect to $\delta_{GR}$ each one of the effective core radii may be enlarged or shrunk depending on the magnitude and the sign of the nonminimal coupling parameter. However, let us notice that in order to ensure $\delta \in \mathbb{R}$ the radicand in (\ref{dmain}) must obey $\tilde{\alpha}>-1/20$.

On the other hand, for $M_{c}$ we also obtain  two solutions, each of them associated with the two core radii found above. By writing it in terms of the corresponding GR expression, $M_{GR}=-\frac{16\pi \eta}{3\sqrt{\lambda}}$, we shall have 
\begin{equation}
\label{mcore} 
M_{c}=-\left| M_{GR} \right | \left\{ \left[\frac{1}{2}+\frac{\epsilon}{2}\sqrt{1+20\tilde{\alpha}} \right]^{3/2}-\frac{15\tilde{\alpha}}{\sqrt{\frac{1}{2}+\frac{\epsilon}{2}\sqrt{1+20\tilde{\alpha}}}}\right\}.
\end{equation}

\begin{figure}[b!]
\begin{center}
{\includegraphics[width=10cm,height=7cm]{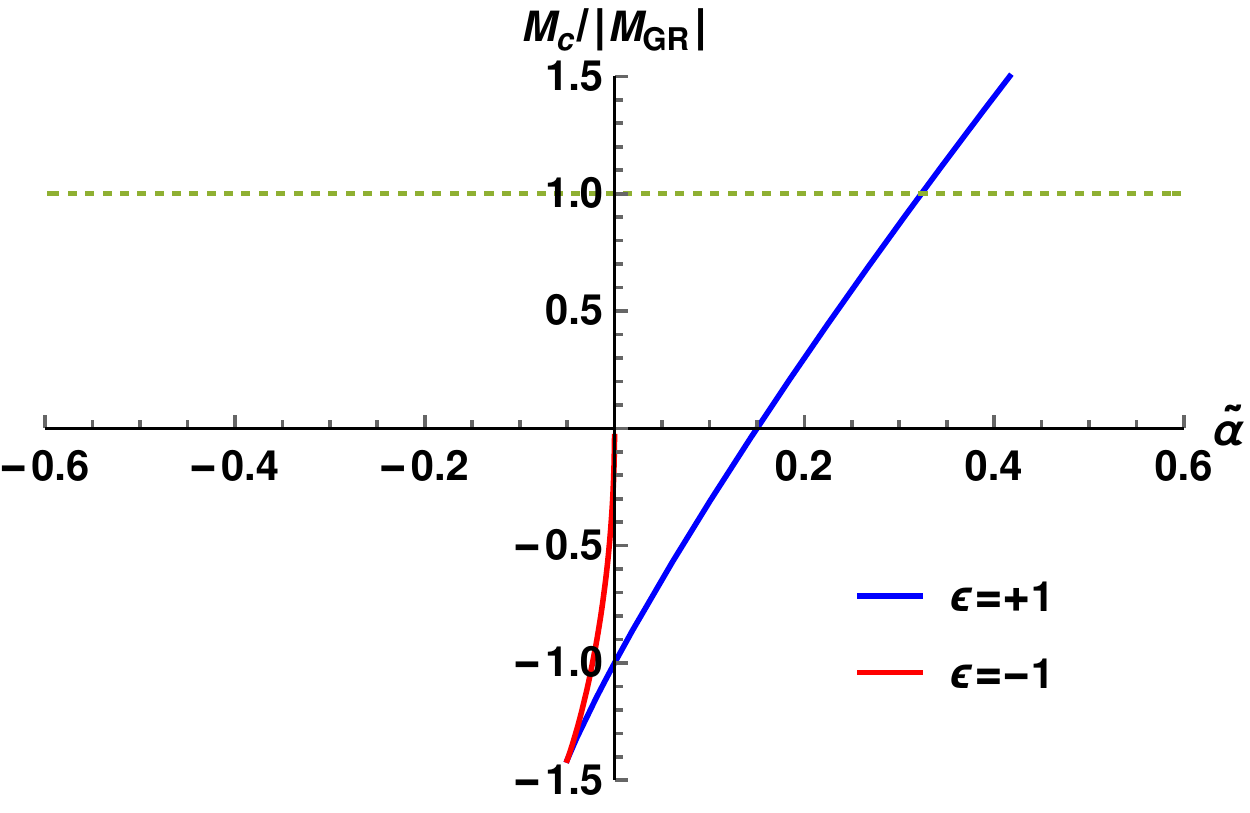}}
\end{center}
\caption{Variation of the ratio $M_{c}/| M_{GR}|$ with the parameter $\tilde{\alpha}$ for both $\epsilon=+1$ and $\epsilon=-1$ cases. The range of $\tilde{\alpha}$ is restricted to $|\tilde{\alpha}|<1$ in order to respect the weak coupling approximation. For the case $\epsilon=-1$, the mass is always negative (blue curve). Whereas, for $\epsilon=+1$, the nonminimal coupling parameter may render positive mass to the global monopole for $\tilde{\alpha}>3/20= 0.15$. The dashed line represents the increased/decreased threshold for the mass of the nonminimal global monopole compared with the GR one.} 
\end{figure} 

\subsection{Discussion}
It is shown in \cite{lousto} that the same steps followed above using the GR (exterior and interior) solutions yield a negative monopole mass, which is easily confirmed by taking $\tilde{\alpha}=0$ in (\ref{mcore}). The main consequence of that is an emergence of a repulsive gravitational potential around the monopole. This peculiar character of the monopole mass was verified  by the authors, not only by an analytical approach based on ({\ref{HLmodel}), but also through numerical techniques. In the present work, we shall concentrate on this same analytical procedure to examine this issue. A more complete analysis of the model requires the employment of a numerical treatment, which we shall leave for a future work.   

the Fig. 1 shows the profile of the ratio $M_{c}/| M_{\textrm{\tiny GR}}|$ as a function of the nonminimal matter-curvature coupling, $\tilde{\alpha}$. The cases $\epsilon=+1$ and $\epsilon=-1$ correspond to the red and blue curves, respectively. We focus on small values of $\tilde{\alpha}$, in agreement with the weak coupling assumption, so that our analysis encompasses the range $|\tilde{\alpha}|<1$. Nevertheless, it is clear by looking at these two curves that the mass $M_{c}$ is not defined to all $\tilde{\alpha}\in \mathbb{R}$ in that range anyway. As we have previously pointed out for the case of $\delta$, from the radicand in (\ref{mcore}) it is possible to set a condition for $M_{c} \in \mathbb{R}$, which implies in a restriction on the domain of the function $\frac{M_{c}}{| M_{GR}|} (\tilde{\alpha})$ to $\left\{\tilde{\alpha}\in \mathbb{R}\; |\; \tilde{\alpha}>-1/20= -0.05\right\}$. As the plot shows us, in both $\epsilon$ cases the  effective value of the monopole mass is quite sensitive to the intensity of the nonminimal coupling. The dashed line sets a threshold above (below) which the mass increases (decreases) in comparison with its GR counterpart as an effect coming directly from the nonminimal matter-curvature exchange. Although the case $\epsilon=-1$ allows for different magnitudes of the mass in comparison with GR, its sign remains negative for all the permitted values of $\tilde{\alpha}$. However, the case $\epsilon=+1$ leads to an interesting result, in which the monopole mass may acquire a positive sign, as long as $\tilde{\alpha}>3/20= 0.15$. This means that, contrary to the GR global monopole, the nonminimal one may exert attractive gravitational force on the material particles moving around it. This is an original contribution of this work. 

%In fact, this result regarding the positive monopole mass fits to the conclusions of \cite{nucamendi}, where a nonminimal coupling between monopole and gravity has been also considered, although through a different coupling term, that is introduced in a purely {\it ad hoc} way.  

%So, the result (\ref{mcore}) tells us that if the global monopole couples nonminimally to gravity at a strong enough magnitude its mass may get positive sign. 
Notice that a positive mass becomes a possibility, even though we had found for the interior solution the same de Sitter--like structure as the standard Hahari-Loust\'o metric. In the context of GR, this de Sitter--like character of the interior solution is usually presented as intrinsically related to the negative sign for the defect's mass since as such it would come up as a repulsive gravity effect. This fact attests to the central role of the here obtained exterior solution in defining the sign of the monopole's mass in our model. 

In Fig. 2, we also plot the variation of  $\delta/| \delta_{GR}|$ with respect to $\tilde{\alpha}$. In the case $\epsilon=-1$, the function $\frac{\delta}{| \delta_{GR}|} (\tilde{\alpha})$ has domain $\left\{\tilde{\alpha}\in \mathbb{R}\; |\; -1/20= -0.05\leq \tilde{\alpha}\leq 0 \right\}$. Moreover, it is clear that, for such a choice of $\epsilon$, all the permitted values of $\tilde{\alpha}$ leads to $\delta/| \delta_{GR}|<1$, meaning that the nonminimal coupling to gravity implies a shrinkage of the core's size. On the other hand, the case $\epsilon=+1$, whose domain is given by  $\left\{\tilde{\alpha}\in \mathbb{R}\; |\; -1/20= -0.05\leq \tilde{\alpha}\leq 0 \right\}$, admits both decrease or increase of the nonminimal monopole's core in comparison with the Barriola-Vilenkin solution. It is worth recalling that the model given by (\ref{HLmodel}) is only a toy model and as such does not show all the aspects present in the global monopole physics in its stringent details, although it is a model that shares most of the key features of the true scenario. 

\begin{figure}[ht]
\begin{center}
{\includegraphics[width=9cm,height=7cm]{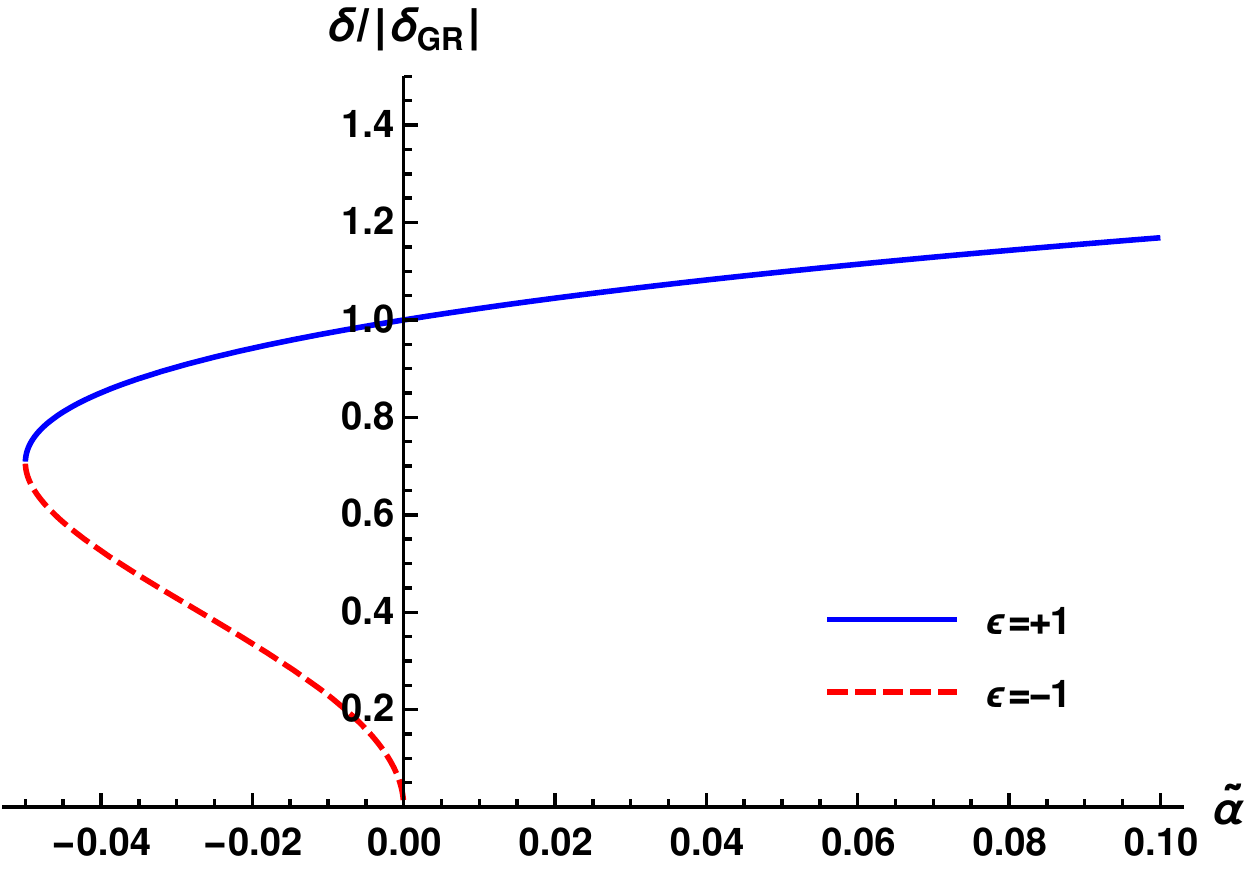}}
\end{center}
\caption{Variation of the ratio $\delta/| \delta_{GR}|$ with the parameter $\tilde{\alpha}$ for both $\epsilon=+1$ and $\epsilon=-1$ cases.} 
\end{figure}

\section{On the black hole with monopole charge}

Henceforth, we shall study a system consisting of a Schwarzschild black hole carrying a global monopole that is nonminimally coupled to the spacetime curvature. In this case, the mass term appearing in the solution is the mass of the black hole itself, so that we shall denote it as $M$, in order to distinguish it from the monopole's mass presented in the previous section.   
%According to Barriola and Vilenkin such a configuration could arise from a black hole that swallows a global monopole, thus acquiring its charge \cite{barriola}.

\subsection{The event horizons}
%\footnote{These singularities are those ones that cannot be removed by means of a mere coordinate change.}

From (\ref{Rscalar1}) we notice that the Ricci scalar diverges for $r=0$. Since $R(r)$ is a scalar quantity, this fact reveals to us the presence of a curvature singularity at the origin, such as is verified in the usual Schwarzschild solution. If the global monopole is treated as an extended object, endowed with a finite core as it was in the previous section, we have seen that one has actually a regularity at origin, instead of a singularity. However, if the monopole shows up as a charge of a black hole of mass $M$, so that the core's mass is $M_{c}<<M$, the singularity at $r=0$ stems from the fact that we are now approximating the defect as a pointlike object, devoid of any inner structure.
%\footnote{In analogy to what happens to a point-like charge located at the origin of the coordinate system, whose eletric field varying as $r^{-2}$ blows up at $r=0$.}. 

In the Schwarzschild case, such a central singularity is surrounded by an event horizon represented by the surface $r_{h}=2GM$. For a Schwarzschild black hole hosting a global monopole inside it, the radius of such a surface becomes $r_{h}=\frac{2GM}{1-\Delta}$.  
In this subsection we intend to analyze how this event horizon radius is affected by the presence of a nonminimal global monopole. 

Before proceeding with this analysis, it is convenient to introduce the new set of dimensionless variables below
\begin{equation}
\label{dim0}
x\equiv r/G M, \;\;\;\;\textrm{and}\;\;\;\sigma\equiv\frac{\alpha}{G^2M^2}\ .
\end{equation}
Furthermore, let us recall that since we are now dealing with a black hole carrying a monopole charge, the $r^{-3}$-term in the exterior metric shall not be neglected as it was in the previous section, in the present case, the weak coupling assumption takes as reference a different length scale, as well as the fact that the mass $M$ is not the mass of the core $M_c$ anymore, but the mass of the black hole itself. These two features enhance the impact of such a term, so that it will be kept in the equations.

Thus, in terms of the variables (\ref{dim0}), the solution encoded in (\ref{solA1}) and (\ref{solB1}) may be rewritten as
\begin{equation}
\label{solA3} 
A(x)^{-1}=1-\Delta - \frac{2}{x}+\frac{14 \sigma \Delta}{x^2}-\frac{18 \sigma \Delta}{x^3},
\end{equation}
and
\begin{equation}
\label{solB3}     
 B(x)=1-\Delta - \frac{2}{x}+\frac{20\sigma \Delta}{x^2}-\frac{30 \sigma \Delta}{x^3}.
\end{equation}
As it is well known, the radial position of the event horizon $r_{h}$ associated with a given static solution may be found through $g^{11}(r_{h})=0$. For the metric functions above, this involves solving the algebraic equation $A(x_{h})^{-1}=0$, whose roots shall give the corresponding horizons radii of the solution. 

%In the Fig.3-a, the influence of negative $\sigma$ is examined. Contrary to GR case, where only one event horizon shows up, we see in such a case that even small deviations from GR lead to the appearance of pairs of horizons: events and Cauchy ones. The plot shows us that by increasing the nonminimal coupling (within the weak coupling assumption), one shifts the pair of horizons farther and farther from the central singularity. On the other hand, in the Fig.3-b, where $\sigma>0$ values are considered, it is shown that only one event horizon appears, exhibiting the same behaviour as the outer ones of the $%\sigma<0$ cases.

An alternative way to depict the occurrence of these horizons is by plotting the graph of the function $A(x)^{-1}$ for different $\sigma$ values, so that we can identify the corresponding zeros and examine how the horizon locations are changed by the nonminimal coupling. This is done in Fig. 3, where the behavior of the metric function $A(x)^{-1}$ is displayed for negative and positive values of $\sigma$. As we did before, in all these curves the solid angle deficit has been fixed as $\Delta=0.3$. 
In both panels, we verify that in the reference case, $\sigma=0$, only one horizon appears at $x_{h}=2/(1-\Delta)$, in accordance with what is expected for the standard Barriola-Vilenkin metric. 

In Fig.3 (a), the influence of a negative $\sigma$ is portrayed. Contrary to the GR case, where only one event horizon shows up, we see here that even small deviations from GR lead to the appearance of pairs of horizons: outer horizons ($x_{+}$) which are the event horizons in the usual sense and the inner ones, the so-called Cauchy horizons ($x_{-}$). The plot shows us that by increasing the nonminimal coupling (within the weak coupling assumption), one shifts this pair of horizons farther and farther from the central singularity. However, it is clear that while the nonminimal coupling exerts a strong impact on the location of the Cauchy horizon, its influence in shaping the event horizon is considerably diminished. This is an already expected result, since the contributions of the $x^{-2}$ and $x^{-3}$ terms in (\ref{solA3}) become significant for small $x$. Furthermore, let us notice that these two surfaces split the spacetime into three distinct regions:
\begin{itemize}
\item i: $x_{+}<x$, where the metric function $A^{-1}(x)$ is positive, so the coordinate $t$ is timelike while $r$ is spacelike;  
\item ii: $x_{-}<x<x_{+}$, where $A^{-1}(x)$ is negative and hence $t$ and $r$ turn into spacelike and timelike, respectively;
\item iii: $x<x_{-}$, where the sign of $A^{-1}(x)$ becomes positive again, so that the coordinates $t$ and $r$ reacquire the original nature they had in the region (i).
\end{itemize}

In this vein, this solution clearly mimics what happens in the Reissner-Nordstr\"om (RN) spacetime, except that in the present case one does not find any naked singularity as long as the nonminimal coupling remains restricted to the weak coupling condition $|\sigma|<1$ that must be indeed obeyed by the here studied metric [given by (\ref{solA3}) and (\ref{solB3})]. As it occurs in the RN case, the presence of this extra inner horizon shall also change the nature of the central singularity, while it is spacelike for $\sigma=0$, it becomes timelike for $|\sigma|\neq0$. 

%The inner and outer horizons become closer and closer each other for increasing values of $\sigma$, up to collapsing to an only one for some value of this parameter around $\sigma \approx 0.16$. If the intensity of the nonminimal coupling is above this value, the black hole becomes horizonless, revealing the arising of naked singularities in such a spacetime.   

\begin{figure}[!t]
  \centering
  \subfloat[]{\includegraphics[width=0.5\textwidth]{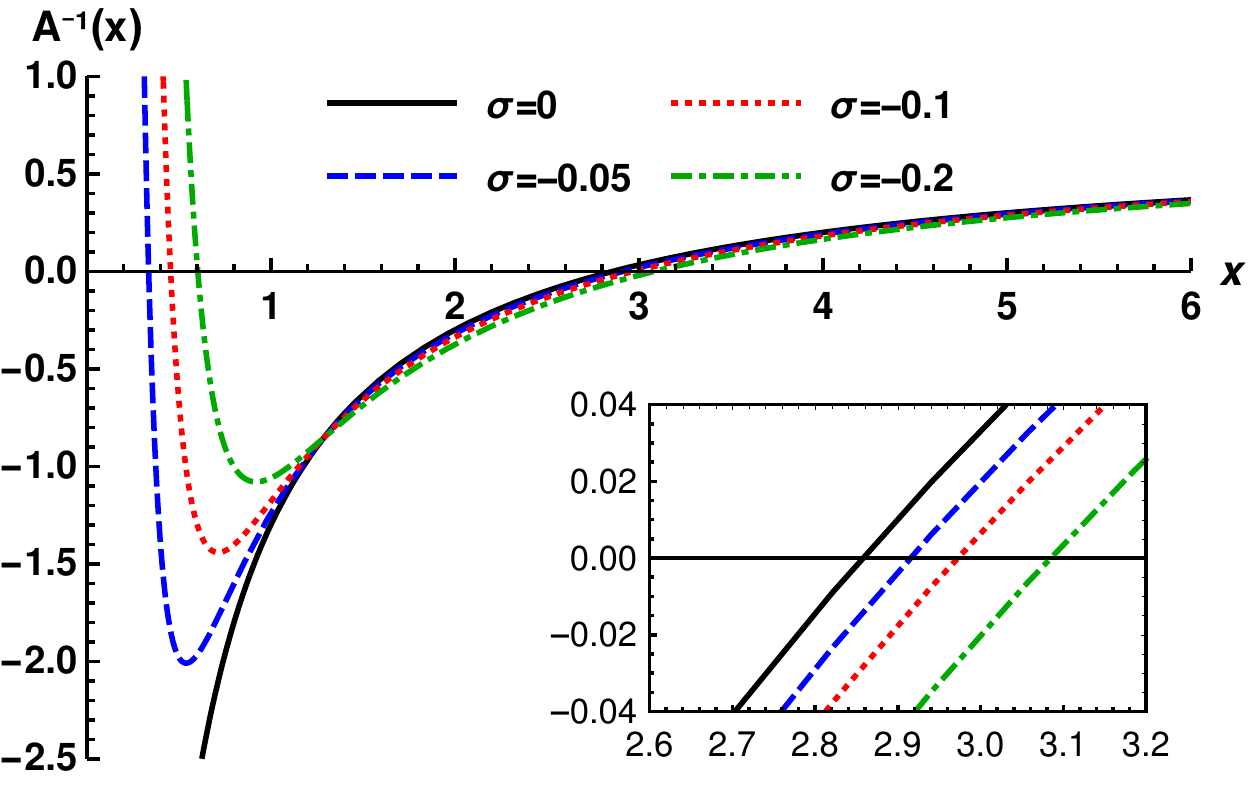}\label{fig:f1}}
  \hfill
  \subfloat[]{\includegraphics[width=0.5\textwidth]{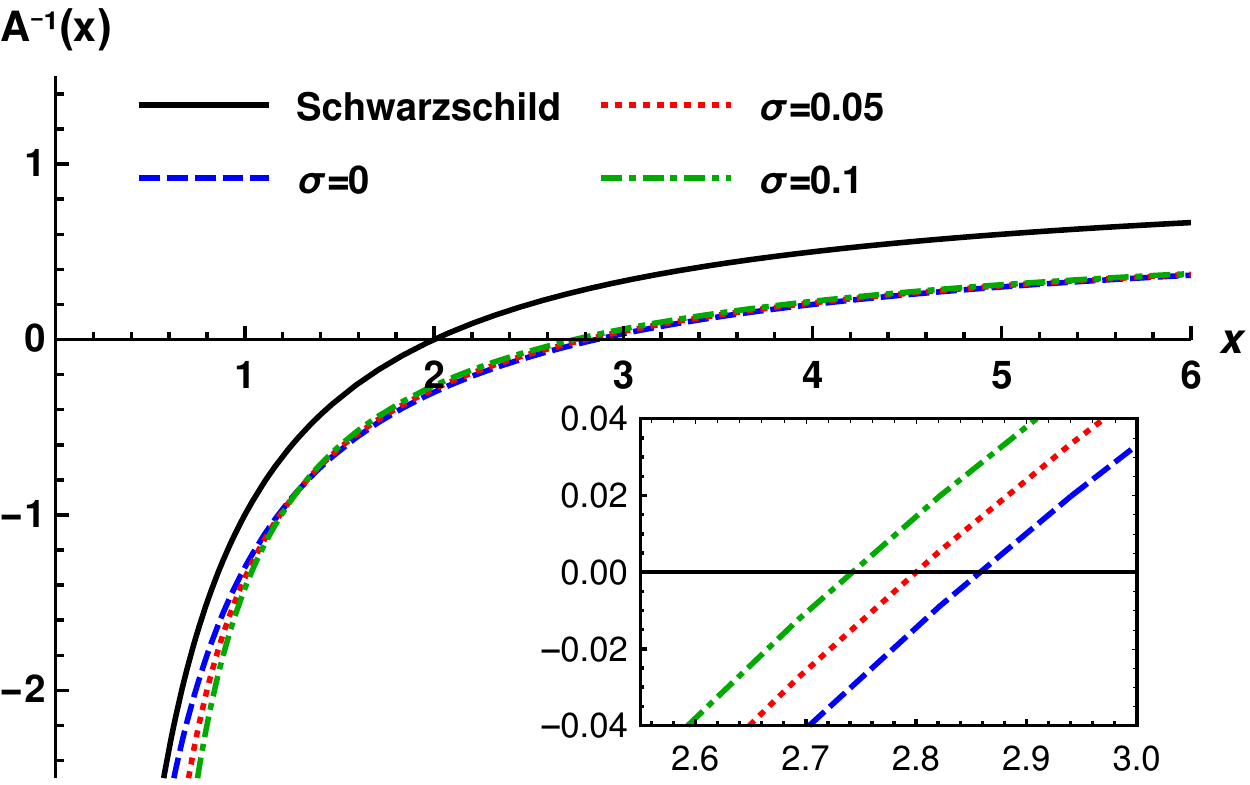}\label{fig:f2}}
  \caption{The metric function $A^{-1}(x)$ for different magnitudes of the nonminimal coupling parameter, considering the illustrative value of $\Delta=0.3$. (a) The $A(x)^{-1}$ curves for the $\sigma<0$ cases. (b) The profile of this metric function for $\sigma>0$ values. The zeros of this function correspond to the locations of the horizons.}
\end{figure}  
%\begin{figure}[ht]
%\begin{center}
%{\includegraphics[width=10cm,height=6cm]{plotA1.pdf}}
%\end{center}
%\caption{The metric function $A^{-1}(x)$ for different $\sigma$ values with $\Delta=0.3$. The zeroes of this function show us the locations of the horizons. Contrary to GR case, where only one event horizon shows up, even small deviations from GR reveals the appearance of pairs of horizons: events and Cauchy ones. The plot shows that increasing values of $|\sigma|$ (within the weak coupling assumption) make the pair of horizons to shift farther and farther from the central singularity.} 
%\end{figure} 
In the context of GR, the appearance of Cauchy horizons in solutions like RN spacetime represents a potential loss of the predictability of the theory. On the other hand, some results indicate instabilities of these horizons against small perturbations involving test massless fields \cite{zamo}, and also a blueshift instability, which means a divergence in the flux of radiation received by an observer crossing this horizon \cite{chand}, leading to a formation of a ``mass-inflation" singularity at such a surface \cite{poisson}. However, other authors show that such instabilities may be circumvented or at least alleviated by considering a RN spacetime plus a cosmological constant \cite{vitor} or assuming that the black hole experiences an accelerated motion \cite{dest}.
In a future study, it would be interesting to revisit these aspects within the context of the solution presented in this work.

As it is exihibited in Fig.3 (b), for $\sigma>0$ we verify a repetition of the same behavior that is seen in the $\sigma=0$ case, in which a single one event horizon forms. However, in contrast to the submodel $\sigma<0$, here the event horizon surfaces become closer and closer to the central singularity as $\sigma$ is increased. Nonetheless, let us notice that it occurs in a quite mild manner, showing that this property is just slightly sensitive to the nonminimal coupling parameter. On the other hand, it is evident that the nontrivial topology encoded in the parameter $\Delta$ has a more decisive role in determining the size of the horizon, as it is demonstrated by the gap that we see between $\Delta=0$ (Schwarzschild) and the remaining $\Delta\neq0$ cases.  

\section{Geodesic motion of test particles}
Now, we will be interested in examining the motion of test particles in the surroundings of such a hairy black hole. It is known that such a geodesic motion in a given geometry may be studied through the Lagrangian ${\cal L}_{g}$ below,
\begin{equation}
\label{Lg}
{\cal L}_g=g_{\mu\nu} \frac{d x^{\mu}}{d \tau} \frac{d x^{\nu}}{d \tau}=k, 
\end{equation}
where $k=0$ or $-1$ denotes massless and massive particles, respectively. In this section we are going to discuss the motion of massive particles around the aforementioned black hole and in this case the affine parameter, $\tau$, will coincide with the proper time for such a particle traveling along timelike geodesics. It is convenient to fix the motion on the equatorial plane $\theta=\frac{\pi}{2}$, which makes (\ref{Lg}) assume the following form:
\begin{equation}
\label{Lg1}
{\cal L}_g= -B(r)\left(\frac{dt}{d\tau}\right)^2+A(r)\left(\frac{dr}{d\tau}\right)^2+r^2 \left(\frac{d\varphi}{d\tau}\right)^2,
\end{equation}
given the metric (\ref{metric}). The Lagrangian above contains clearly two cyclic coordinates, $t$ and $\varphi$, which leads to the arising of the conserved quantities below,
\begin{equation}
\label{ctes}
E\equiv B(r) \frac{d t}{d\tau}\;\;\;\textrm{and}\;\;\;\;\;L\equiv r^2 \frac{d \varphi}{d\tau},
\end{equation}
where $E$ and $L$ mean, respectively, the total energy and angular momentum per unit of mass of the test particle. Using these variables in (\ref{Lg1}) and setting $k=-1$ we get the following orbital equation:  
\begin{equation}
\label{en}
\frac{A(r)B(r)}{2}\left(\frac{d r}{d \tau}\right)^2+V_{\textrm{eff}}(r)={\cal E},
\end{equation}
where ${\cal E}\equiv \frac{E^2}{2}$ and the effective potential being given by 
\begin{equation}
\label{veff}
V_{\textrm{eff}}(r)\equiv\frac{B(r)}{2}\left(\frac{L^2}{r^2}+1\right).  
\end{equation}
Using (\ref{AB}) in (\ref{en}) we may express (\ref{en}) as follows:
\begin{equation}
\label{en1}
\frac{1}{2}e^{\left(\frac{6\alpha \Delta}{r^2}\right)}\left(\frac{d r}{d \tau}\right)^2+V_{\textrm{eff}}(r)={\cal E}.
\end{equation}

It is evident from the equation above that it describes a motion subject to the condition ${\cal E}-V_{\textrm{eff}}(r)>0$. Moreover, the radial positions for which ${\cal E}=V_{\textrm{eff}}(r)$ represent the turning points of the motion. With (\ref{en1}) at hand, we are able to investigate the possible orbits to be followed by material test particles. In particular, we shall be interested in stable circular motion around our hairy black hole. It is well known that, given an effective potential associated with a certain spacetime through Eq. (\ref{en1}), any stable circular orbit must fulfill the following conditions:
\begin{itemize}
 \item $\dot{r}=0$;
 \item $\partial V_{\textrm{eff}}/\partial r=0$;
 \item $\partial^2 V_{\textrm{eff}}/\partial r^2>0$,
\end{itemize}
where the dot means derivative with respect to the proper time. The first condition implies in ${\cal E}=V_{\textrm{eff}}$, what fixes the energy a test particle 
must have at a given orbit. The second condition tells us that for a stable circular orbit the effective potential must be a critical point and it gives us a polynomial equation whose roots indicate the radial positions of the orbits. However, this is a necessary but not sufficient condition for stable circular orbits. It is well known that the nature of these critical points depends on the sign of the second derivative: they will be either minimum or maximum points, depending on the positive or negative sign of the second derivative of $V_{\textrm{eff}}(r)$, respectively. The former shall provide stable orbits, whereas the latter will result in unstable ones. In this vein, the third condition provides the final requirement that stable circular orbits have to obey. According to what we have learned from classical mechanics, in stable orbits the orbiting particle tends to return to its original orbit whenever its motion is slightly perturbed. Whereas for unstable orbits, even a small perturbation is enough to make the particle to depart further from its original orbit.

\subsection{The innermost stable circular orbit}

In the context of GR, stable circular orbits are not possible at arbitrarily small radii, contrary to what is predicted in the Newtonian gravity, where these orbits may exist for all radial positions. In Einstein gravity, considering the standard Schwarzschild spacetime, the minimum radius at which these orbits are possible is $r_{\textrm{\tiny ISCO}}=6GM$ \cite{kaplan,landau,moore}, where ISCO denotes the innermost stable circular orbit. This parameter is useful in the study of accretion disks, as it represents the inner edge of the disk \cite{abramo} and its value is usually affected by intrinsical properties of the central mass around which the orbit takes place, for instance, the spin of the black hole or the equation of state of the neutron star \cite{torok,shun}. It is possible to show that an orbiting particle at the ISCO has angular momentum and energy given by $L_{\textrm{\tiny{ISCO}}}=2\sqrt{3}G M$ and $E_{\textrm{\tiny{ISCO}}}=\sqrt{8/9}$. In GR, the occurrence of the ISCO takes place at a intermediate condition between $\partial^2 V_{\textrm{eff}}/\partial r^2>0$ and $\partial^2 V_{\textrm{eff}}/\partial r^2<0$, in which one has the so-called marginally stable orbit, where $\partial^2 V_{\textrm{eff}}/\partial r^2=0$.

All these ISCO parameters are slightly modified if the Schwarzschild black hole under analysis is endowed with a global monopole charge. For this case, the black hole mass is rewritten as $M\rightarrow M/(1-\Delta)$, thus changing the quantities $r_{\textrm{\tiny{ISCO}}}$ and $L_{\textrm{\tiny{ISCO}}}$, whereas $E$ turns out to be $E=\sqrt{(1- \Delta)8/9}$ \cite{dadhich}. 
%Let us recall that in the present work, we wish to assess how much some immediate physical properties of a gravitating global monopole are affected when it couples nonminimally to its gravitational field. 
%{\bf Since we wish to inspect how a given orbital motion is affected by the nonminimal matter-curvature coupling, it is helpful to use the scaled ISCO parameters above as a reference in our study, as they already contain the monopole's charge term. Therewith, we ensure that any new effect appearing in the motion under consideration is originated purely from the nonminimal coupling.} 
So, it is helpful to use the scaled ISCO parameters above as a starting point of our study about how the nonminimal matter-curvature coupling influences a given orbital motion.

In addition to the variables (\ref{dim0}), let us introduce the redefinition $l \equiv L/GM$. For convenience, we will express the results of this section in terms of these dimensionless variables. So, at leading order in $\Delta$ and $\alpha$ the effective potential will be expressed in terms of such variables as
\begin{equation}
\label{Veffx}
\tilde{V}_{\textrm{eff}}(x)=\frac{1}{2} \left\{1-\Delta-\frac{2}{x}+\frac{2[10\sigma \Delta + 6(1+\Delta)]}{x^2}-\frac{3[10\sigma \Delta + 8(1+2\Delta)]}{x^3}+\frac{240\sigma \Delta}{x^4}-\frac{360\sigma \Delta}{x^5}\right\},  
\end{equation}
%%%%%
where we have used
\begin{equation}
\label{lisco}
l^2=\frac{12}{(1-\Delta)},
\end{equation}
when going from (\ref{veff}}) to (\ref{Veffx}), which corresponds to the $l$ at ISCO, rescaled by the global monopole term.

\subsection{Discussion}

In Fig. 4, the behavior of the effective potential (\ref{Veffx}) and its first derivative as functions of $x$ is shown, for different values of the nonminimal coupling parameter. In Fig. 4 (a) one compares the $\tilde{V}_{\textrm{eff}}(x)$ for different cases. The green curve is the effective potential associated with the usual Schwarzschild black hole, for which $\Delta=0$ and $\sigma=0$, with the black dot indicating the ISCO position at $x=6$. For the sake of illustration, for the remaining cases the global monopole parameter is fixed as $\Delta=0.3$. The black curve is the Barriola-Vilenkin black hole, which corresponds to the Schwarzschild metric with a standard global monopole charge. From this plot we verify that increasing values of $\sigma$ give rise to maximum points in the $\tilde{V}_{\textrm{eff}}(x)$ curve corresponding to higher and higher values of this function. This signals the appearance of a potential barrier that an arbitrary test particle traveling from the infinity has to overcome in order to reach the central singularity. So, the strength of the nonminimal coupling contributes to increase more and more of such a barrier. Additionally, one also verifies that the size of the gap between the green curve and the bundle of curves below is related to the difference between $\Delta=0$ (assumed at the green curve) and $\Delta=0.3$ (valid at the black, blue, and red ones). So, the larger the parameter $\Delta$ is, the wider will be this gap. 

\begin{figure}[!b]
  \centering
  \subfloat[]
  {\includegraphics[width=8.5cm,height=6.5cm]{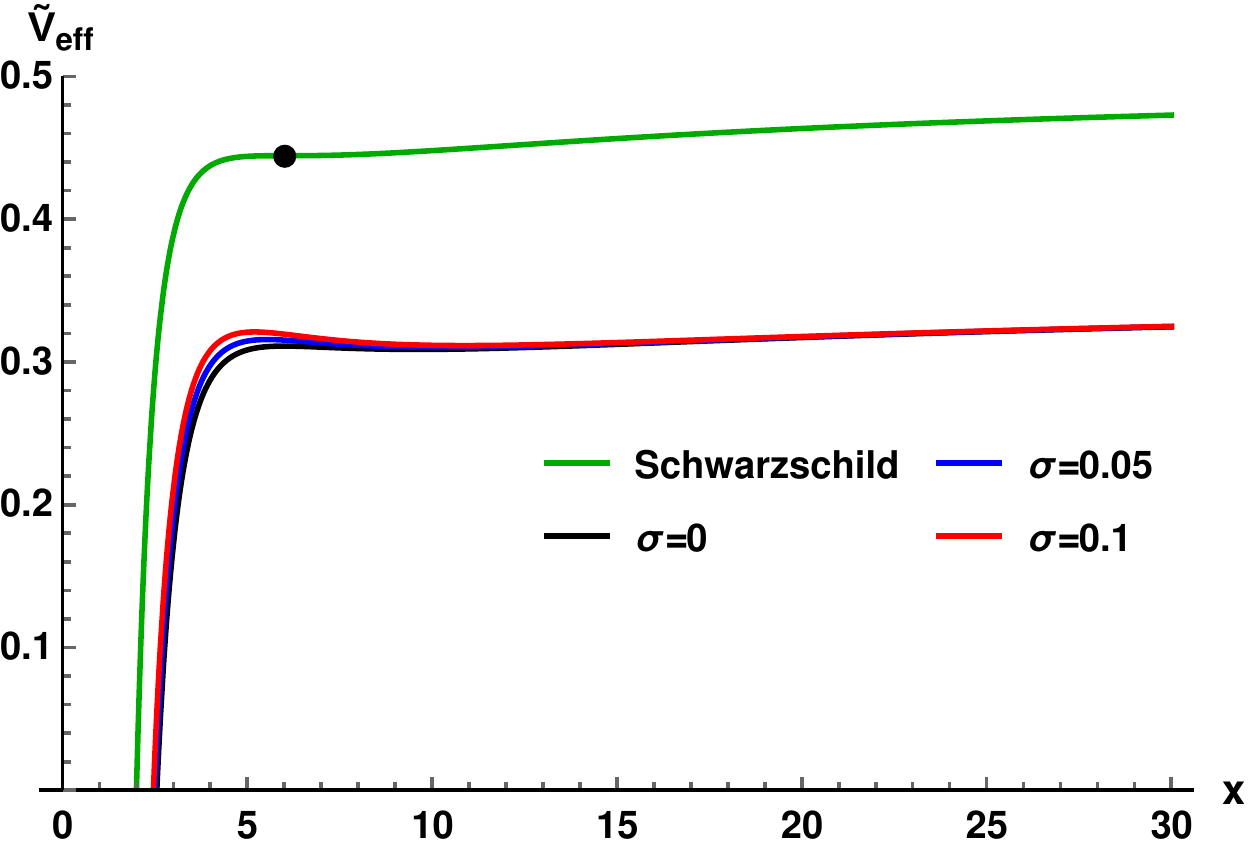}\label{fig:f0}}
  \hfill
  \subfloat[]
{\includegraphics[width=9cm,height=6 cm]{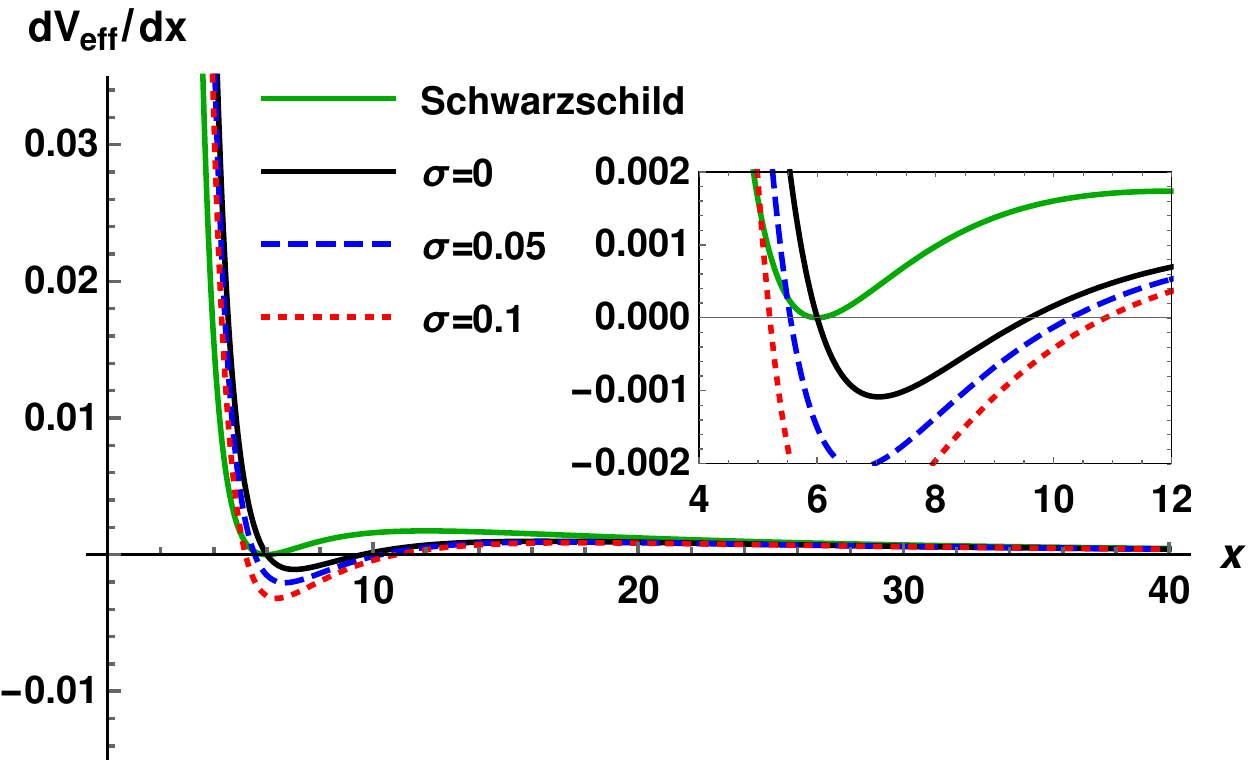}}

\caption{Behavior of the effective potential with respect to the reparametrized distance $x$. (a) Choices in which $\Delta>0$ make the $\tilde{V}_{\textrm{eff}}$ curves to be shifted downward, while the nonminimal coupling parameter gives rise to new critical points for $\tilde{V}_{\textrm{eff}}$, both maxima and minima. (b) Makes clear such a new critical points' appearance by showing the profile of the first derivative of $\tilde{V}_{\textrm{eff}}$.}
  \end{figure}

The Fig. 4 (b) allows us to locate the position of possible either unstable or stable orbits, by examining the position of the critical points of the effective potential, i.e the zeroes of equation $d\tilde{V}_{\textrm{eff}}/dx$. This plot help us to undestand more clearly the nature of the critical points. Notice that in Schwarzschild case the derivative of the effective potential has only one zero in a point in which there is no change of sign for $d\tilde{V}_{\textrm{eff}}/dx$. This aspect characterizes what is known by marginally stable orbit existing at the ISCO radius. On the other hand, for the remaining non-Schwarzschild cases, we verify changes of sign for the function  $d\tilde{V}_{\textrm{eff}}/dx$ when its respective zeroes are reached. For those sign changes in which one goes from positive (negative)  to negative (positive) sign, we have the presence of a maximum (minimum) point there.          

Looking closely at the effective potential curves one sees this emergence of either new maxima and minima points for each $\sigma$'s value in a more evident way. In Fig. 5 (a) the maxima points for different intensities of the nonminimal coupling are shown. There, we notice that growing values of $\sigma$ lead to the appearance of possible unstable orbits closer and closer to the center of the black hole. On the other hand, Fig. 5 (b) shows that the larger $\sigma$ is, the farther is the position at which the new minima will arise. These minima will correspond to the ISCO radii for this hairy black hole and their occurence will have consequences on the matter accretion around such compact objects. As we have mentioned before, the ISCO radius encloses the inner border of the accretion disk by separating the surface of the disk itself from the plunging region where the matter falls almost freely (with negligible torque) into the central black hole. It is known that the energy flux and the temperature, as well as the luminosity of an accretion disk depend on the ISCO position, so the results depicted in Fig. 5 constitute a potential observational signature for this black hole studied here \cite{santiago,harko}. This issue will deserve proper attention from us in the future.   

We finish our analysis by examining the $\sigma<0$ cases. In Fig. 6, we represent the curves for the effective potential as well as its first derivative in order to discuss its capability in engendering stable orbits, as we did for the previous case. The first difference that is immediately realized between the present case and the previous one is the formation of an infinite ``centrifugal" barrier on the left. That barrier shall prevent a test particle coming from the infinity with the angular momentum (\ref{lisco}) to reach the central singularity, regardless of the energy it has. 

In Fig. 6(a), in addition to the $\tilde{V}_{\textrm{eff}}(x)$ profile, we also use two dots to indicate the positions of both the Cauchy and event horizons for the respective curves, %In fact, from the indicated position of the horizons, it is clear that even the Cauchy horizon shall be inaccessible to this particle.
with which it is possible to see that for the analyzed $\sigma$ values, the $\tilde{V}_{\textrm{eff}}(x)$ has minima located between both horizons. In principle, this would point the possibility of the existence of stable orbits in this region; however, we have to rule them out of our model since, as mentioned in the previous section, when a particle traverses the event horizon, its radial coordinate turns into a timelike one. As such, it will inevitably decrease toward the central singularity, in a progressive infall that only finishes when the particle crosses the Cauchy horizon. So, this reinforces the fact that, despite what the $\tilde{V}_{\textrm{eff}}$ curves tell us, such a intermediary region is not suitable for the existence of stable orbits. \footnote{Nevertheless, it is worth mentioning the interesting Ref. \cite{Doku}, which discusses the viability of orbits (even for the development of life) in the innermost region, inside the Cauchy horizon of charged or rotating black holes.}

On the other hand, the right panel illustrates the possible appearance of both stable and unstable orbits in the region outside the event horizon, as the zeros of the function $d\tilde{V}_{\textrm{eff}}/dx$ show us. We can see that, for $\sigma=-0.05$, we will have an unstable orbit at the point where the derivative changes its sign from ``$+$" to ``$-$" (maximum point) and an outer stable one at the point where the opposite sign change occurs (minimum point). This second orbit will correspond to the ISCO one for the case $\sigma=-0.05$, located at $x_{\textrm{ISCO}}\approx 8.62$. It is instructive to compare this result with the $x_{\textrm{ISCO}}\approx 10.30$ obtained for $\sigma=0.05$, as it is portrayed in the blue curve of the Fig. 5 (b). This evinces the crucial role of the $\sigma$ sign in defining how the orbital motion will be structured. Moreover, as demonstrated by the evolution of the curves with growing values of $\sigma$, when the nonminimal coupling is increased, such orbits in the region external to the horizon cease to exist. 

\begin{figure}[!t]
  \centering
  \subfloat[]{\includegraphics[width=0.5\textwidth]{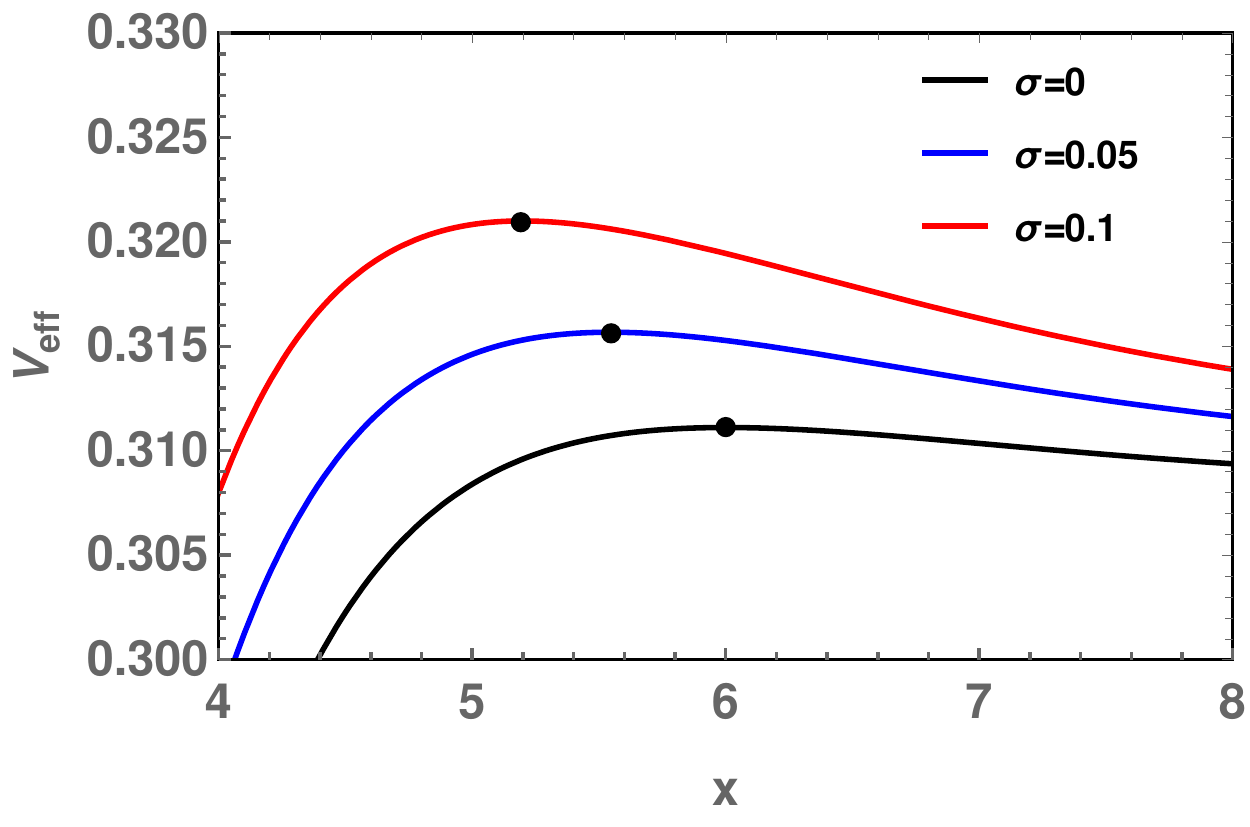}\label{fig:f1}}
  \hfill
  \subfloat[]{\includegraphics[width=0.5\textwidth]{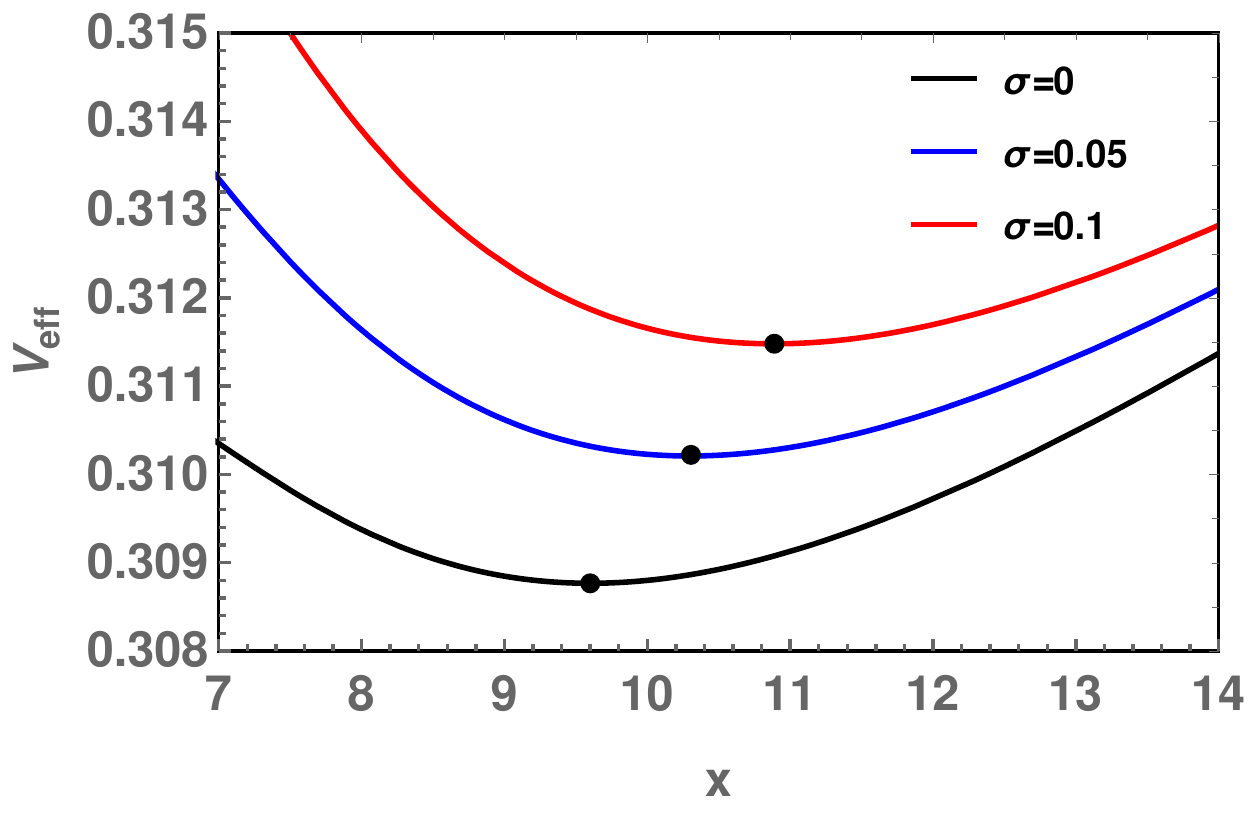}\label{fig:f2}}
  \caption{The arising of new possible orbits for a material particle subject to (\ref{Veffx}). (a) The left panel shows how the nonminimal coupling dictates the appearance of unstable orbits around the central black hole. (b) On the other hand, the right panel illustrates the emergence of stable orbits for different values of $\sigma$.}
\end{figure}  
%%%%%%%
\begin{figure}[!t]
  \centering
  \subfloat[]{\includegraphics[width=0.5\textwidth]{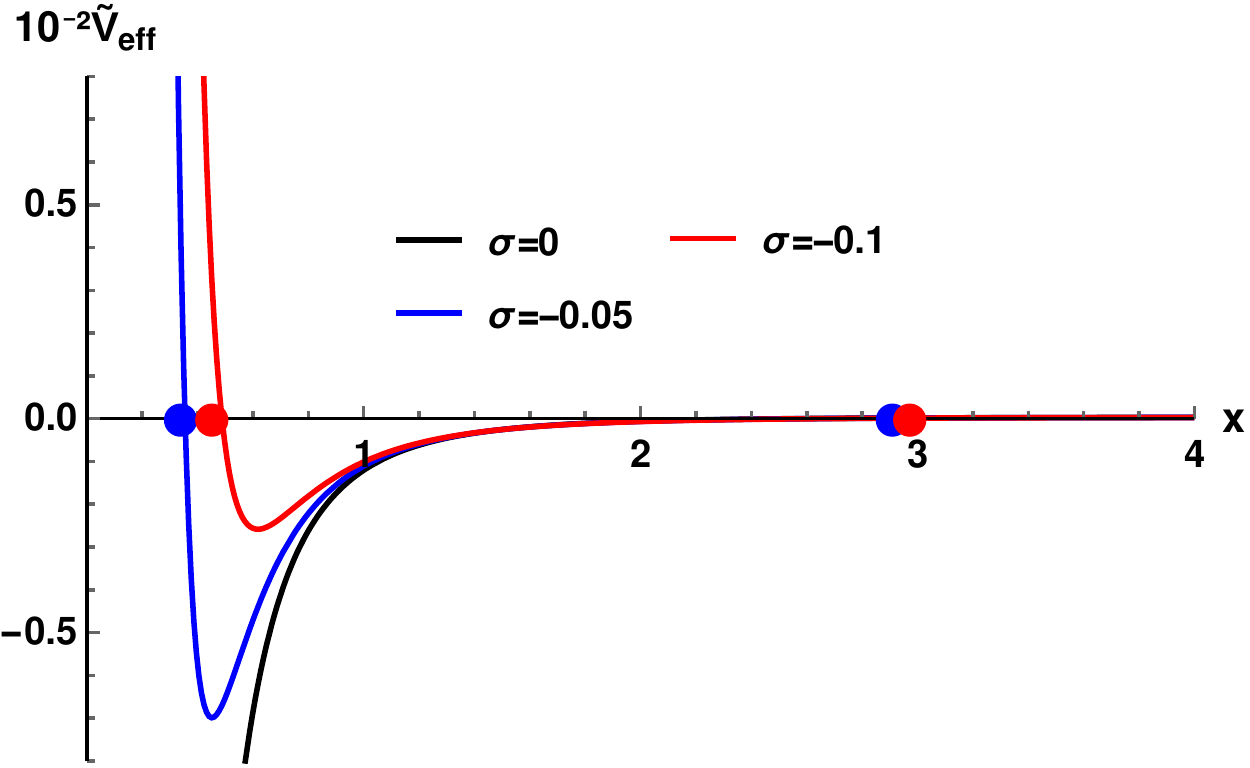}\label{fig:f1}}
  \hfill
  \subfloat[]{\includegraphics[width=0.5\textwidth]{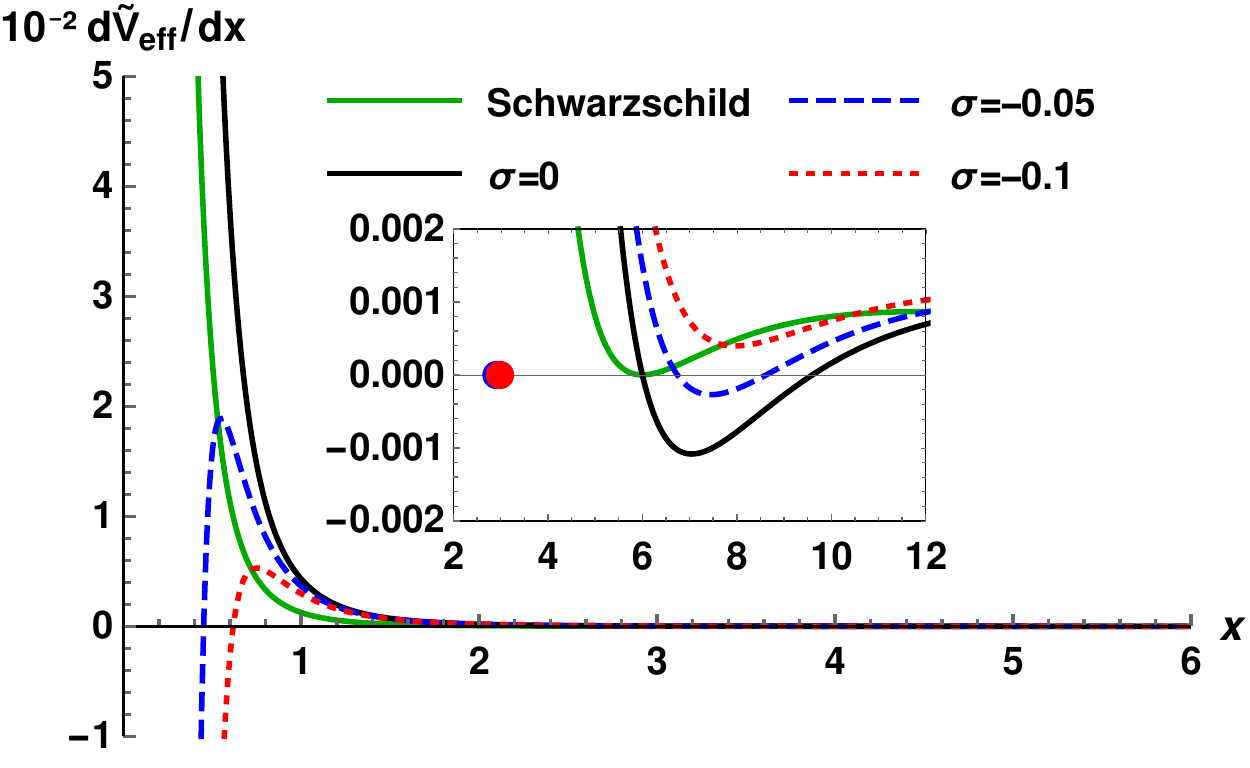}\label{fig:f2}}
  \caption{Behavior of the effective potential $\tilde{V}_{\textrm{eff}}$ for the $\sigma<0$ submodel. The dots in both panels indicate the location of the horizons (Cauchy and event ones). (a) The arising of an infinite barrier for the effective potencial felt by a particle endowed with the angular momentum here analyzed. (b) Displays the critical points' appearance by showing the profile of the first derivative of $\tilde{V}_{\textrm{eff}}$.}
\end{figure}  

\section{The gravitational light bending}
At last, we shall discuss the effects such a black hole may exert on light rays traversing its surroundings. As we shall see, there will be combined effects coming from both the nonminimal matter-curvature coupling and the nontrivial topology induced by the global monopole. The orbital equation obtained from the null geodesic associated with the background (\ref{metric}) is found by taking $k=0$ in (\ref{Lg}), which gives
\begin{equation}
\label{geodL} 
 \frac{e^{\left(\frac{6\alpha \Delta}{r^2}\right)}}{r^4}\left(\frac{dr}{d\varphi}\right)^2+\frac{B(r)}{r^2}=\frac{1}{b^2},
\end{equation}
where $b\equiv L/E$ is the impact parameter of the light ray trajectory. As we are working within the weak coupling framework, it is convenient to expand the exponential term in (\ref{geodL}) for a small $\alpha$, so that this equation now becomes
\begin{equation}
\label{geodL1} 
 \frac{1}{r^4}\left(1+\frac{6\alpha \Delta}{r^2}\right)\left(\frac{dr}{d\varphi}\right)^2+\frac{B(r)}{r^2}=\frac{1}{b^2}.
\end{equation}
This equation allows us to obtain the characteristic length of the light bending physics, which is the so-called closest approach distance of the light ray with respect to the central mass, denoted by $r_0$. In Newtonian gravity, one does not assume any gravitational light deflection, so that the quantities $b$ and $r_{0}$ actually coincide. However, in the context of GR, $r_{0}$ deviates from $b$ by means of relativistic corrections, typically proportional to $GM$ powers. In an extended gravity, like the one we are handling, one expects that new corrections appear on the expression for $r_{0}$. By definition, $r_0$ is the distance at which the light experiences a turning point in its motion, so $\frac{dr}{d\varphi}=0$. By imposing this condition in (\ref{geodL}) we find following algebraic equation for $r_0$
\begin{equation}
\label{r0} 
r_0^{5}-b^2(1-\Delta)r_0^{3} -2GMb^2 r_0^{2}+20\alpha \Delta b^2 r_{0}-30GM\alpha \Delta=0.
\end{equation}
Fortunately, for our purposes, it is not necessary to solve exactly the above quintic equation. We may instead obtain a tractable expression for $r_0$ by solving (\ref{r0}) to the leading order in $GM$ and $\alpha$. As we shall see in this section, taking these two quantities as parameters in a perturbative expansion reveals to be fully compatible with the premises we will assume in our study of the light bending. Solving (\ref{r0}) along these lines, one finds the following approximated solution:
\begin{equation}
\label{solr0}
r_{0}\backsimeq b\sqrt{1-\Delta}-\frac{GM}{1-\Delta}+\frac{10\alpha \Delta}{b(1-\Delta)^{3/2}}.
\end{equation}
For reasons that will become clear later, we shall be interested in the quantity $1/r_{0}$, whose expression can be obtained from (\ref{solr0}),
\begin{equation}
\label{r0ap}
\frac{1}{r_{0}}\backsimeq \frac{1}{b\sqrt{1-\Delta}}+\frac{GM}{b^2(1-\Delta)^2}-\frac{10\alpha \Delta}{b^3(1-\Delta)^{5/2}}.
\end{equation}
It is usual to redefine the radial coordinate as $u=r^{-1}$. In terms of such variable, the equation of motion (\ref{geodL1}) can be recast as follows:
\begin{equation}
\label{eqU}
(1+6\Delta \alpha u^2)\left(\frac{du}{d\varphi}\right)^2+u^2(1-\Delta)-2GM u^3+20\alpha \Delta u^4-30GM\alpha \Delta u^5=\frac{1}{b^2}.
\end{equation}

An important consideration about the angle variable $\varphi$ is in order. It is assumed that the black hole we are studying here carries a global monopole charge, possibly after having swallowed it, as we have mentioned previously. Notice that, even without such interaction, the global monopole may affect the geodesic motion of light particles in its spacetime due to the solid deficit angle that emerges around it from the parameter $\Delta$. Hence, it is useful to redefine the variable $\varphi$ so that the residual influence of the solid deficit angle is taken into account. This leads us to introduce a new angle variable $\bar{\varphi}$, which comes from the original $\varphi$ as follows:
\begin{equation}
\label{def} 
\sqrt{1-\Delta}\varphi \longrightarrow \bar{\varphi},
\end{equation}
which can be infered directly from (\ref{BV-metric}), considering $\theta$ at the equatorial plane, $\theta=\frac{\pi}{2}$. Such a redefinition for the angle variable due to the presence of a solid deficit angle has already been adopted in the literature, as one can see in Ref. \cite{banerjee}. In terms of this new variable, we will have  $u(\varphi(\bar{\varphi}))=\bar{u}(\bar{\varphi})$ and the Eq. (\ref{eqU}) turns out to be
\begin{equation}
\label{eqU1}
(1+6\Delta \alpha \bar{u}^2)(1-\Delta)\left(\frac{d\bar{u}}{d\bar{\varphi}}\right)^2+\bar{u}^2(1-\Delta)-2GM \bar{u}^3+20\alpha \Delta \bar{u}^4-30GM\alpha \Delta \bar{u}^5=\frac{1}{b^2}.
\end{equation}
We can obtain a second order differential equation from (\ref{eqU1}) by taking its derivative with respect to $\bar{\varphi}$,
\begin{equation}
\label{eqU2}
2(1+6\Delta \alpha \bar{u}^2)(1-\Delta)\left(\frac{d^2\bar{u}}{d\bar{\varphi}^2}\right)+12\alpha \Delta \bar{u}(1-\Delta)\left(\frac{d\bar{u}}{d\bar{\varphi}}\right)^2+2\bar{u}(1-\Delta)-6GM\bar{u}^2+80\alpha\Delta\bar{u}^3-150GM\alpha \Delta \bar{u}^4=0.
\end{equation}

As it is well known, an orbital equation like (\ref{eqU2}) may be solved by means of a perturbative treatment, in which we consider only up to first order accuracy in $GM$ and $\alpha$. In this procedure, the function $\bar{u}(\bar{\varphi})$ is split into its zeroth and first order parts as follows:
\begin{equation}
\label{barU}
\bar{u}=\bar{u}_{0}+\bar{u}_{1}.
\end{equation}
In our approach, we will consider that the global monopole contributes both to the zeroth and first perturbative orders. While its solid angle deficit is treated as a background property, its nonminimal coupling to gravity is assumed as a first order effect. 
That is why we take the global monopole parameter $
\Delta$ as a background quantity, of the same order as $\bar{u}_{0}$. With this choice we intend to look more closely at the effects coming from the nonminimal coupling by enhancing the influence of the $\alpha$ parameter on the gravitational light bending.

Using (\ref{barU}) in (\ref{eqU2}) one finds the zeroth and first order solutions for the respective differential equations, so that the general solution up to first order in $GM$ and $\alpha$ reads
\begin{equation}
\label{solU}
\bar{u}(\bar{\varphi})=\frac{\textrm{sin}\bar{\varphi}}{R_0}+\frac{3\alpha\Delta (9+\Delta)}{2R_{0}^3(1-\Delta)}\bar{\varphi}\textrm{cos}\bar{\varphi}-\frac{\alpha \Delta (7+3\Delta)}{8R_{0}^3(1-\Delta)}\textrm{sin}(3\bar{\varphi})+\frac{GM}{2R_{0}^2(1-\Delta)}\textrm{cos}(2\bar{\varphi})+a_{0}\textrm{sin}\bar{\varphi}+\frac{3GM}{2R_{0}^2(1-\Delta)},
\end{equation}
where $R_{0}$ and $a_0$ are integration constants to be determined with the aid of the initial conditions 
\begin{equation}
\bar{u}(\bar{\varphi}=\pi/2)=1/r_0\;\;\;\;\;\textrm{and}\;\;\;\;\;\frac{d\bar{u}}{d\bar{\varphi}}|_{\bar{\varphi}=\pi/2}=0.
\end{equation}
Using them in (\ref{solU}) and then comparing it with (\ref{r0}) we find the expressions for $R_0$ and $a_0$, which are
\begin{equation}
R_{0}=b\sqrt{1-\Delta}\;\;\;\textrm{and}\;\;\;a_0=-\frac{\alpha \Delta(87+3\Delta)}{8b^3(1-\Delta)^{5/2}}. 
\end{equation}
So, the final form of the general solution to the leading order in $GM$ and $\alpha$ gives
\begin{eqnarray}
\label{genU}
\bar{u}(\bar{\varphi})=\left[\frac{1}{b\sqrt{1-\Delta}}-\frac{\alpha \Delta(87+3\Delta)}{8b^3(1-\Delta)^{5/2}}\right]\textrm{sin}\bar{\varphi}&+&\frac{3\alpha\Delta (9+\Delta)}{2b^3(1-\Delta)^{3/2}}\bar{\varphi}\textrm{cos}\bar{\varphi}-\frac{\alpha \Delta (7+3\Delta)}{8b^3(1-\Delta)^{5/2}}\textrm{sin}(3\bar{\varphi})\nonumber\\&+&\frac{GM}{2b^2(1-\Delta)^2}\textrm{cos}(2\bar{\varphi})+\frac{3GM}{2b^2(1-\Delta)^2}.
\end{eqnarray}

Now, we are ready to compute the magnitude of the light deflection caused by this black hole. The approach to be used to obtain this result depends crucially on the nature of the spacetime under consideration. For asymptotically flat geometries, like the Schwarzschild one, it is possible to assume the light coming from a far source located at infinity, so that the light trajectory is initially a straight line that is gradually bent as the light approaches the surroundings of the spherical distribution of mass. After the light reaches the distance of closest approach with respect to the central mass, it is deflected and then continues its motion toward an observer at infinity, thus resuming progressively its former aspect of a straight line trajectory. The deflection angle is defined as the angle between the two asymptotes that encompass the full path of the light, before and after it is deflected by the central mass. As it is well known, for a Schwarzschild metric, the magnitude of such an angle is given to the leading order in $GM$ by $\delta_{\textrm{def}}=\frac{4GM}{c^2b}$.  
However, notice that the spacetime given by the metric (\ref{metric}) is not asymptotically flat in the usual sense, which recommends us to adopt an alternative procedure to study the gravitational light deflection. For this purpose, we will make use of the method proposed by Rindler and Ishak \cite{rindler}. In that approach, the authors analyze the light bending associated with a Schwarzschild--de Sitter (SdS)  spacetime by introducing the variable $\Psi$, which is the angle between the radial direction and the light trajectory at a certain point $P(r,r(\varphi))$. Given the metric (\ref{metric}) one may verify that $\Psi$ must obey the following relation:
\begin{equation}
\label{rind}
\textrm{tan} \Psi=r\sqrt{[A(r)]^{-1}}\left|\frac{dr}{d\varphi}\right|^{-1}.
\end{equation}
In Ref. \cite{arakida}, the reader may find a very intuitive and purely geometrical way to understand the relation above. Additionally, as discussed in \cite{bhadra}, we can also use the orbital equation (\ref{geodL}) to rewrite (\ref{rind}) in a more convenient form, namely,
\begin{equation}
\label{rind1} 
\textrm{tan} \bar{\Psi} =\frac{1}{\sqrt{\frac{r^2 B(r_0)}{r_0^2 B(r)}-1}},  
\end{equation}
where, similar to what we did for $\varphi$, we have incorporated the deficit angle property in a new definition for $\Psi$ given by $\Psi \longrightarrow \bar{\Psi}=\sqrt{1-\Delta}\Psi$. For $r \gg r_0$, one expects the angles $\bar{\Psi}$ and $\bar{\varphi}$ to be very small, which implies $ \textrm{tan} \bar{\Psi} \sim \bar{\Psi}$ as well as $\textrm{sin}\bar{\varphi}\sim \bar{\varphi}$ and $\textrm{cos}(2\varphi) \sim 1$. So, using this assumption in (\ref{genU}) and (\ref{rind1}) we are left with the following expressions for $\bar{\Psi}$ and $\bar{\varphi}$, to the leading order in $GM$ and $\alpha$:
\begin{equation}
\bar{\Psi}=\frac{r_0}{r}-\frac{GMr_0}{r^2(1-\Delta)}+\frac{10\alpha \Delta r_0}{r^3(1-\Delta)}+\frac{GM}{r(1-\Delta)}-\frac{10\alpha \Delta}{rr_{0}(1-\Delta)},
\end{equation}
and
\begin{equation}
\bar{\varphi}=\frac{r_0}{r}+\frac{GM}{r(1-\Delta)}-\frac{10\alpha \Delta}{rr_{0}(1-\Delta)}-\frac{2GM}{r_0(1-\Delta)},
\end{equation}
respectively.

As it is well known, the strength of the gravitational light deflection may be assessed by the one-sided bending angle, $\epsilon$, which consists of the angle between the tangent to the light trajectory at the point $(r,\varphi)$ and the polar axis. For our case, this angle is\footnote{Based on the definitions adopted by Rindler and Ishak \cite{rindler}.} 
\begin{equation}
\label{bend}
\varepsilon \equiv \bar{\Psi}-\bar{\varphi} =\frac{2GM}{r_0 (1-\Delta)}-\frac{GM r_0}{r^2(1-\Delta)}+\frac{10\alpha \Delta r_{0}}{r^3(1-\Delta)}.  
\end{equation}

Let us recall that the distance of closest approach $r_{0}$ is coordinate dependent and not a measurable quantity. For this reason, it is suitable to express the bending angle in terms of the apparent impact parameter $b$, which is in fact a coordinate independent variable. So, using (\ref{solr0}) and (\ref{r0}) in the equation above we can rewrite it as
\begin{equation}
\label{nend1} \varepsilon =  \bar{\Psi}-\bar{\varphi} =\frac{2GM}{b (1-\Delta)^{3/2}}-\frac{GM b}{r^2\sqrt{1-\Delta}}+\frac{10\alpha \Delta b}{r^3\sqrt{1-\Delta}}.  
\end{equation}

In calculations of the bending angle in the context of asymptotically flat geometries, one usually considers observer and source located at infinity by taking the limit $r\rightarrow\infty$ in the orbit equation. This assumption, however, is not proper within nonasymptotically flat metrics. For example, the SdS spacetime is endowed with the de Sitter horizon at a finite radial distance $r\approx\sqrt{3\Lambda}$, making the procedure of taking the limit $r\rightarrow\infty$ in the orbit equation totally meaningless. In fact, since all observed stars and galaxies are located at a finite distance from us, any realistic scenario should take into account the role of finite-distance corrections on the bending angle expression. In this work, we are interested in assessing the influence of such contributions. So, following the procedure adopted in Refs. \cite{bhadra,pacheco} we may rewrite the result above in terms of the positions both of the source ($r_{\textrm{s}}$,$\varphi_{\textrm{s}}$) and the observer ($r_{\textrm{obs}}$,$\varphi_{\textrm{obs}}$) with respect to the central mass. So, when the contributions at source and observer positions are considered, the total bending is
\begin{equation}
\label{tot-epsi}
\epsilon_{\textrm{total}}=(\bar{\Psi}_{\textrm{s}}+\bar{\Psi}_{\textrm{obs}})-(\bar{\varphi}_{\textrm{s}}+\bar{\varphi}_{\textrm{obs}}),
\end{equation}
from which we will find the explicit form of the total deflection angle, $\epsilon_{\textrm{total}}$, 
\begin{equation}
\label{total-epsi} 
 \epsilon_{\textrm{total}}=\frac{4GM}{b (1-\Delta)^{3/2}}-\frac{GM b}{\sqrt{1-\Delta}}\left(\frac{1}{r_{\textrm{s}}^2}+\frac{1}{r_{\textrm{obs}}^2}\right)+\frac{10\alpha \Delta b}{\sqrt{1-\Delta}}\left(\frac{1}{r_{\textrm{s}}^3}+\frac{1}{r_{\textrm{obs}}^3}\right).
\end{equation}
The first term of (\ref{total-epsi}) is promptly identified as the usual GR result for the deflection angle caused by a spherical distribution of mass, rescaled by the presence of the monopole charge through the term $(1-\Delta)^{3/2}$. Whereas, the second and third terms on the right-hand side of (\ref{nend1}) correspond to finite-distance corrections that shall be important to the light bending phenomenon only if the finite positions of the source and observer with respect to the central mass need to come into play. 

One may also note that for $\Delta=0$ Eq. (\ref{total-epsi}) coincides with the Eq. $(25)$ of \cite{pacheco} for $\Lambda=0$, as it is expected. In Ref. \cite{ishihara}, using the Gauss-Bonnet theorem, the authors examine the impact of finite-distance corrections on the bending of light for two nonasymptotically flat backgrounds, namely, the SdS spacetime as well as an exact solution of the Weyl conformal gravity. 

Finite-distance effects on the bending of light shall be significant in situations where the source is near enough, such as in planned astrometric missions like the Laser Astrometric Test of Relativity mission \cite{turyshev}, designed to probe the deflection of light by the solar gravity by means of laser interferometry between two microspacecrafts whose lines of sight pass close by the Sun.   

\subsection{Estimating the contribution of the nonminimal monopole term to the deflection angle}

In Ref. \cite{suzuki} is analyzed the observational consequences of such finite-distance terms on the gravitational lensing caused by both the Sun and Sagittarius A* (Sgr A*). In that paper, the authors estimate how much such corrections affect the bending of light for these two gravitating systems.

Let us follow a similar route by considering the hypothesis in which the Sgr A* black hole at the Galactic Center carries a nonminimal global monopole charge. In this case, we will be interested in making a rough estimate about the contribution of the third term of (\ref{total-epsi}), here denoted by $\delta \epsilon$, to the deflection angle. For this purpose, we assume that the light coming from a given source gets deflected by Sgr A* and reaches an observer at Earth. With this premise, it becomes reasonable to neglect the term $1/r_{\textrm{obs}}^3$, since $r_{\textrm{obs}}$ is far larger than the impact parameter, while a source star may be in the bulge of the Galaxy. For this reason, we may consider that the most relevant finite-distance contribution comes from the term proportional to $1/r_{\textrm{s}}^3$, so that
\begin{equation}
\label{epsi}
\delta \epsilon \sim \frac{10 \alpha \Delta b}{\sqrt{1-\Delta}\; r_{\textrm{s}}^3}.
\end{equation}
For a typical grand unification scale one has $\Delta=8\pi G \eta^2 \sim 10^{-5}$ \cite{vilenkin}, which puts $\sqrt{1-\Delta}$ quite close to unity. Moreover, let us suppose that the light source is the star S2, which was tracked during its May 2018 closest approach to Sgr A* \cite{s2}. The observations showed that such a pericenter approach was given by $r_{s}=1400\;GM$, in which the mass of the central black hole is $M=4\times10^6M_{\odot}$. Within the weak deflection limit, where the impact parameter $b>>GM$, we shall assume $b=10^3GM$. Additionally, let us rewrite (\ref{epsi}) in terms of the dimensionless parameter $\sigma$ introduced in (\ref{dim0}). So, we are left with   
\begin{equation}
\label{epsi1}
\delta \epsilon \sim 10\sigma \left(\frac{8 \pi G \eta^2}{10^{-5}}\right) \left(\frac{b}{10^3\;GM}\right)\left(\frac{1400\;GM}{r_{s}}\right)^{3} \;\;\mu \textrm{arcsec}.
\end{equation}
As it is expected, the impact of such a correction on the deflection angle depends crucially on the strength of the nonminimal coupling. Since we are adopting the weak coupling hypothesis, the values of the parameter $\sigma$ will be restricted to $\sigma < 1$. If the model is marginally inside the weak coupling regime with $\sigma\sim10^{-1}$, the corresponding $\delta \epsilon$ would be of the order of $1\;\mu \textrm{arcsec}$, which is $1$ order of magnitude beyond the current sensitivity achieved by the Event Horizon Telescope \cite{EHT1,EHT2,EHT3,EHT4,EHT5,EHT6}. However, the improvement of this accuracy up to a submicroarcsecond level is among the prospects of astrometric space missions over the next decades \cite{gaia}. 

On one hand, considering that the enhancement of $\delta \epsilon$ as seen above is greatly favored by the special conditions created by the advent of the closest approach of the star S2, one might expect that the observational signature yielded by the nonminimal coupling of the global monopole to gravity will be, in general, beyond the sensitivity of the present observational techniques for the majority of the cases of stars orbiting Sgr A*. On the other hand, $\delta \epsilon$ shows up as a better observational signature in comparison with the light deflection caused by a standard global monopole, minimally coupled to gravity, for which the corrections on the bending appear through the square and cubic roots of $(1-\Delta)$, whose effect on the bending angle is negligible. 

%\begin{equation}
%\label{dim0}
%x\equiv r/G M; \;\;\;\;\textrm{and}\;\;\;\sigma\equiv\frac{\alpha}{G^2M^2}\ .
%\end{equation}      

%Since, it is reasonable to neglect $\$  let us rewrite this contribution as follows
%\begin{equation}
%\label{epsi}
%\delta \epsilon 10\frac{\alpha \Delta b}{\sqrt{1-\Delta}}
%\end{equation}

\section{Concluding Remarks}

In this work we have studied a gravitating global monopole that couples nonminimally to gravity. We analyzed several consequences of this nonminimal coupling on the study of this topological defect and compared our findings with the known results for the standard Barriola-Vilenkin monopole. Considering a specific gravitational action by which such a matter-curvature interplay is modeled, we managed to find an analytical solution for the respective field equations describing the model in the zone outside the monopole's core. Next, we assumed this nonminimal coupling is small enough, so that all the nonlinear contributions in the coupling parameter were neglected. This assumption provided a significant simplification for the problem, allowing us to reach a clearer interpretation of the results. Then, by assuming a global monopole endowed with a inner structure, we have provided the interior solution of the system within the weak coupling regime. By using the Hahari-Loust\'o analytical model  along with proper matching conditions, we were able to determine some of the main properties of this inner structure, namely, the core radius and the mass contained in it, which were expressed in terms of the parameter of the theory. We have analyzed the range of values of this parameter capable to ensure physically acceptable values for the radius of the core. Additionally, we have shown that, contrary to what is seen in GR, the nonminimal coupling with gravity may provide a positive mass to the global monopole, although the interior metric obeys the same de Sitter--like structure as the Barriola-Vilenkin monopole.  
 
In the second half of our study, we focused on some features of a hypothetical nonrotating black hole carrying a nonminimal global monopole charge. The first aspect we examined was the potential appearence of both event and Cauchy horizons for this spacetime induced by nonzero coupling parameters. In the aftermath, we finished this work by studying the geodesic motion both of time- and lightlike particles in this scenario. For material particles, we discuss the requirements for the existence of stable circular orbits, showing the possible arising of both stable and unstable circular orbits. In the analysis of the light bending, we computed the deflection angle, by expressing it in terms of the finite positions of the source and observer. This hypothesis has a purpose of giving a more realistic character to the study of the light deflection, since all potential sources in the Universe are effectively located at a finite distance from the receiver. On the other hand, we verified that such assumption is crucial to enhance the effect of the nonminimal matter-curvature coupling on the light bending phenomenon. 

In addition to the original contribution of this work {\it per se}, we believe it gives rise also to promising perspectives to be explored in forthcoming opportunities. The collection of new results here obtained reveals an interesting global monopole model which is, furthermore, free of the negative mass plaguing the standard Barriola-Vilenkin framework. However, this model has been constructed by appealing to the outside-the-core approximation, which exempted us from using Eq. (\ref{phi}) along with the field equations, which would form a system of nonlinear coupled differential equations, only solvable by numerical techniques. So, only a full numerical treatment of this set of field equations might confirm if the positiveness of the mass is ensured within such a more general version of this model. 

It is also expected that the emergence of event and Cauchy horizons of different radii in the spacetime of the here investigated hairy black hole may bring consequences for the thermodynamics of this configuration, since all relevant quantities of the black hole thermodynamics are affected by the size of the event horizon radius, as it is well known. Likewise, this may impact the analysis of quantum phenomena that can take place near the event horizon, thus motivating future studies on this issue within the framework of a semiclassical gravity. Furthermore, the arising of Cauchy horizons in such a geometry suggests that studies focused on the analysis of stability of such a surfaces against small perturbations may be a interesting possibility for future efforts.

Moreover, the novel features arising in the investigation of the geodesic motion indicates potential observational prospects, whether for time- or lightlike particles. For the material particles motion, our study pointed out possible signatures to be seen in the accretion disks phenomenon by means of changes in the $r_{\textrm{ISCO}}$ position induced by the nonminimal global monopole. While for the light geodesics, the extra term in the deflection angle related to the nonminimal coupling corresponds to an $r^{-3}$ dependence upon the source and observer positions, which may be distinguished from the remaining two terms in measurements with high enough precision. In this vein, we have provided an order-of-magnitude estimate considering SgrA* as the central black hole responsible for the light deflection, taking as the light source the star S2 in its most recent pericenter approach. 

The light bending study has demonstrated that the model under scrutiny seems to indicate that the property of nonminimal coupling between matter and gravity may perhaps contribute for the increasing of the detectability of topological defects. At which extent would this be confirmed by a detailed analysis involving other observables on an astrophysical scale? The answer for this intriguing question raised by the present work deserves to be pursued in future investigations considering both other cosmic defects and even other models of nonminimally coupled gravity.  

\begin{acknowledgments}
I thank colleagues Eugenio R. B. de Mello, Julio M. Hoff da Silva, Julio C. Fabris and Anderson A. Tomaz for the careful reading of this manuscript and for making important suggestions that helped me to improve considerably this work.
%JMHS thanks to CNPq (Grant no. 303561/2018-1) for partial financial support.
\end{acknowledgments}

\end{document}